\documentclass[aps,twocolumn,superscriptaddress,prl,floatfix]{revtex4}
\usepackage{amssymb}
\usepackage{amsmath}
\usepackage{mathtools}
\usepackage{graphicx}
\usepackage{xfrac}
\usepackage[dvipsnames]{xcolor}
\usepackage{gensymb}
\usepackage{bm}

\def\Idc{I_{\mathrm{DC}}}
\def\Ic{I_{\mathrm{C}}}

\begin{document}
\title{Josephson Detection of Time Reversal Symmetry Broken Superconductivity in SnTe Nanowires}

\author
{C. J. Trimble,$^{1 \dag}$ M. T. Wei,$^{1 \dag}$ N. F. Q. Yuan,$^{2}$ S. S. Kalantre,$^{1}$\\
P. Liu,$^{3, 4}$ H.-J Han,$^{3, 4}$ M.-G. Han,$^{5}$ Y. Zhu,$^{5}$ J. J. Cha,$^{3, 4, 6}$ L. Fu,$^{2}$ J. R. Williams$^{1\ast}$
\\
\vspace*{0.25cm}
\small{$^{1}$\emph{Joint Quantum Institute and Quantum Materials Center, Department of Physics, University of Maryland, College Park, MD USA}}\\
\small{$^{2}$\emph{Department of Physics, Massachusetts Institute of Technology, Cambridge, MA USA}}\\
\small{$^{3}$\emph{Department of Mechanical Engineering and Materials Science, Yale University, New Haven, CT, USA}}\\
\small{$^{4}$\emph{Energy Sciences Institute, Yale West Campus, West Haven, CT, USA}}\\
\small{$^{5}$\emph{Department of Condensed Matter Physics and Materials Science, Brookhaven National Laboratory, Upton, NY, USA}}\\
\small{$^{6}$\emph{Canadian Institute for Advanced Research Azrieli Global Scholar, Toronto, ON, Canada}}\\
\footnotesize{$^\dag$\emph{These authors equally contributed.}}\\
\footnotesize{$^\ast$\emph{To whom correspondence should be addressed; E-mail:  jwilliams@physics.umd.edu.}}\\
}

\date{\today}

\maketitle

\textbf{Exotic superconductors, such as high T$_C$, topological, and heavy-fermion superconductors, require phase sensitive measurements to determine the underlying pairing. Here we investigate the proximity-induced superconductivity in nanowires of SnTe, where an $s\pm is^{\prime}$ superconducting state is produced that lacks the time-reversal and valley-exchange symmetry of the parent SnTe. This effect, in conjunction with a ferroelectric distortion of the lattice at low temperatures, results in a marked alteration of the properties of Josephson junctions fabricated using SnTe nanowires. This work establishes the existence of a ferroelectric transition in SnTe nanowires and elucidates the role of ferroelectric domain walls on the flow of supercurrent through SnTe weak links. We detail two unique characteristics of these junctions: an asymmetric critical current in the DC Josephson effect and a prominent second harmonic in the AC Josephson effect. Each reveals the broken time-reversal symmetry in the junction.  The novel $s\pm is^{\prime}$ superconductivity and the new Josephson effects can be used to investigate fractional vortices~\cite{Chen10, Tanaka18}, topological superconductivity in multiband materials~\cite{Qi10, Hosur14, Guguchia17}, and new types of Josephson-based devices in proximity-induced multiband and ferroelectric superconductors~\cite{Tanaka15, Yerin17}.}

A $s \pm$ superconducting state can arise in the presence of competition between the proximity effect and a repulsive interaction between the effective two bands used in the description of the electronic structure of SnTe~\cite{Hsieh12}. The phase-dependent part of the free energy derived is of the form~\cite{Supp}
\begin{equation}
F(\theta_1, \theta_2)=J\text{cos}(\theta_1-\theta_2)+J_1^{\prime}\text{cos}(\theta_1)+J_2^{\prime}\text{cos}(\theta_2)
\end{equation}

\begin{figure}[t!]
\center \label{fig1}
\includegraphics[width=3 in]{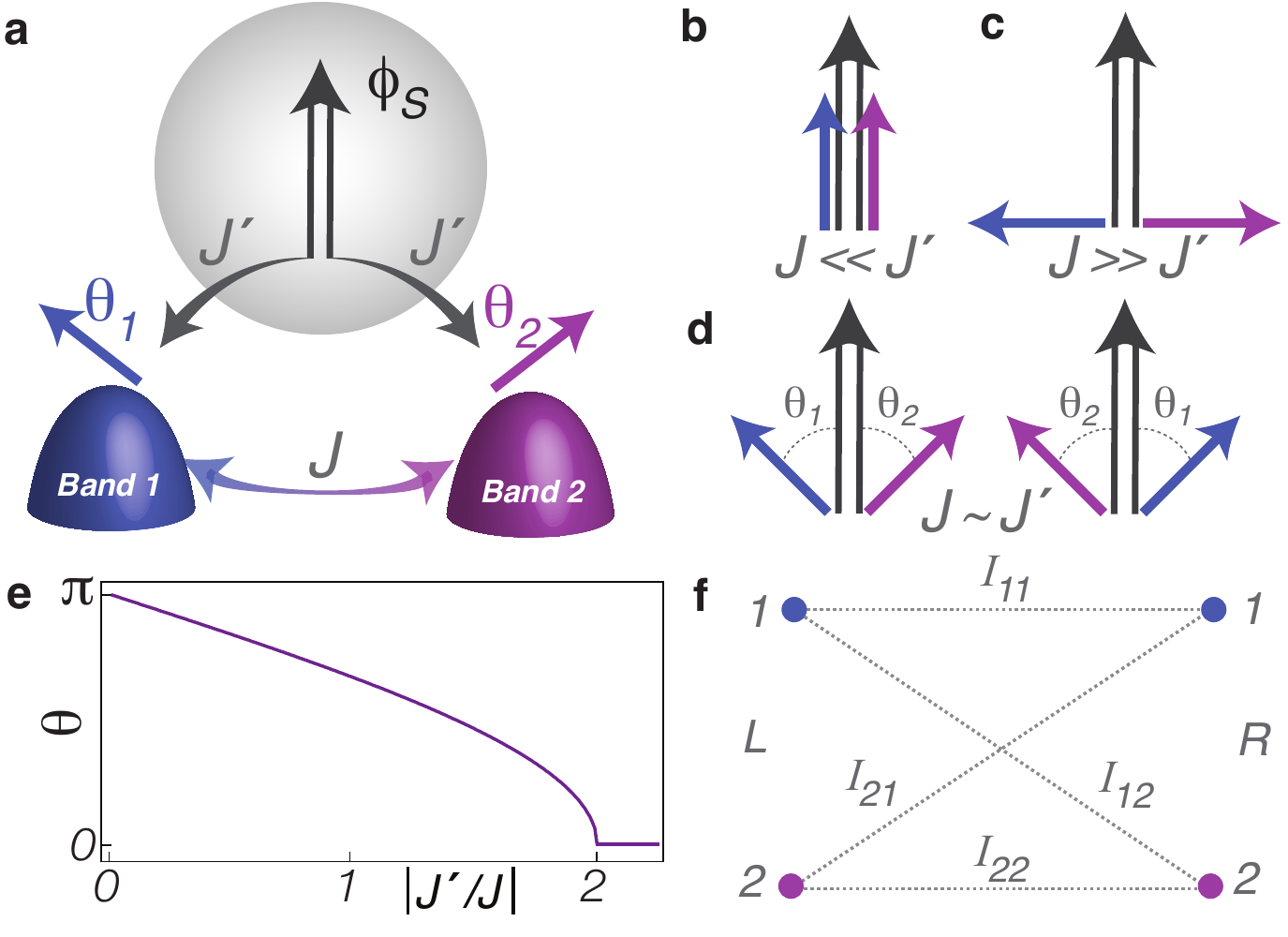}
\caption{\footnotesize{\textbf{Induced $s\pm is^{\prime}$ superconductivity in SnTe. } \textbf{a}, The two bands in SnTe are coupled to the order parameter $\phi_S$ in aluminum via an external pairing field for $J_1^{\prime}=J_2^{\prime}=J^{\prime}$. The interband coupling $J$ is facilitated via the Umklapp process. $\theta_1$ and $\theta_2$ are the phases of individual order parameters in the two bands. \textbf{b}-\textbf{d}, The competition between the coupling strengths $J$ and $J^{\prime}$ results in different relative phases between two bands:  \textbf{b}, When $J\ll J^{\prime}$, the phases tend to align with each other. \textbf{c}, When $J\gg J^{\prime}$, the phases of two bands are out of phase by $\pi$. \textbf{d}, In the intermediate regime  $J\sim J'$, the phases are canted. The two degenerate states in the TRSB case are shown. \textbf{e}, The phase difference between two bands $\theta \equiv \theta_1 - \theta_2$ as a function of the coupling strength ratio $|J^{\prime}/J|$~\cite{Supp}. The nonzero canting angle yields the $s\pm is^{\prime}$ superconductivity in SnTe. \textbf{f}, The four-channel supercurrent flow between two superconducting electrodes $L$ and $R$. The total supercurrent is governed by the phase difference between two conventional superconductors $\phi = \phi_s^R-\phi_s^L$, resulting in a relative rotation that changes the relative amount of supercurrent contributed by each channel. }}
\end{figure}

\begin{figure*}[t!]
\center \label{fig2}
\includegraphics[width=5 in]{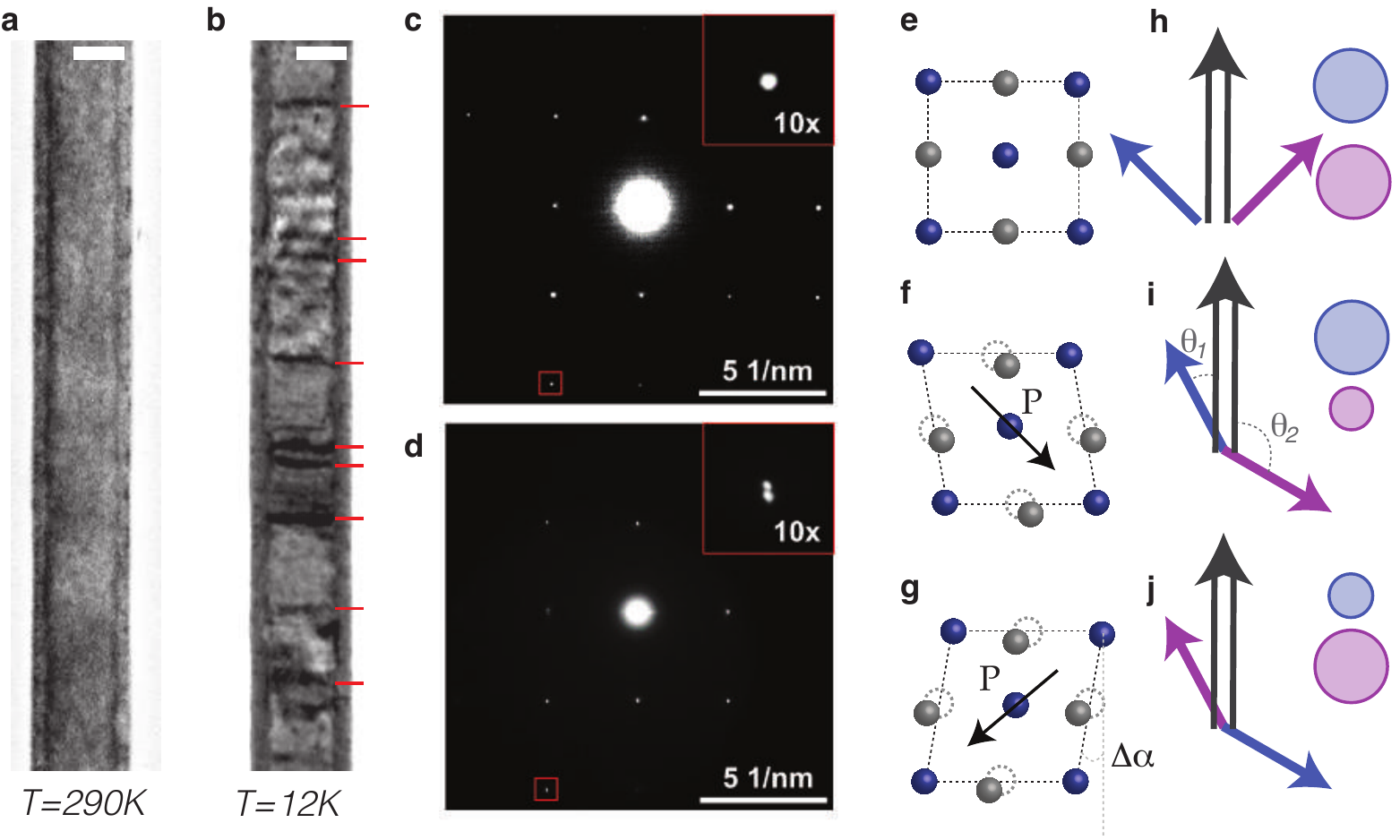}
\caption{\footnotesize{\textbf{Ferroelectric distortion in SnTe nanowires.} \textbf{a-b}, TEM images of a SnTe nanowire at T=290K (\textbf{a}) and T=12K (\textbf{b}). The scale bar in the upper right corner of each image is 50 nm. Dark bands perpendicular to the growth direction (indicated by red lines in \textbf{b}) are domain walls separating different polarization directions. Scale bar in the upper right corner of each image are 50nm. \textbf{c-d}, Diffraction pattern at T=290K (\textbf{c}) and T=12K (\textbf{d}). Splitting of the low-temperature diffraction pattern arises from the two ferroelectric domains in the sample. \textbf{e-g}, The cubic lattice (\textbf{e}) at room temperature undergoes a transition to a rhombohedral lattice at T=80K with two domains (\textbf{f-g}). \textbf{h}, Equal-phase-angle configuration, produced when the density of states of each band at the Fermi energy (indicated in blue and purple) are equal. \textbf{i-j}, Unequal phase angles for the two ferroelectric domains and the parameter $\theta_0$. In \textbf{h-j}, both states which produce a minimum in the free energy are shown.}}
\end{figure*}

\noindent where $\theta_{j=1,2}$ are the phases of the superconducting order parameter in each band (measured with respect to the phase of the proximal aluminum superconductor $\phi_S$), $J^{\prime}$ is a measure of the interband coupling, and $J_1^{\prime}, J_2^{\prime}$ are measures of the external pairing field (provided by the aluminum superconducting leads) to each band, as illustrated in Fig. 1a. Minimization of this free energy dictates that the ground state can allow for a pairing phase difference between the two bands, which depends on the relative strength of $J, J_1^{\prime}$ and $J_2^{\prime}$. If a finite phase difference between bands occurs, the superconducting order parameter for SnTe becomes $s\pm is^{\prime}$: one pocket has an order parameter $\Delta_1 + i \Delta_2$, the other $\Delta_1 - i \Delta_2$, where $\Delta_{j=1,2}$ are the superconducting amplitude on each band.

The free energy is distinct from the conventional free energy of Josephson junctions (JJs). In the ground state, both time-reversal symmetry ($\theta_i \rightarrow - \theta_i$) and valley-exchange symmetry $\theta_1 \leftrightarrow \theta_2$ arising from the four-fold rotational symmetry -- two symmetries, which were preserved prior to inducing superconductivity -- are broken, while their product ($\theta_i \rightarrow - \theta_j$) is preserved. Finally, the competition between $J$ and $J_1, J_2$ should be noted: $J_1$ and $J_2$ want to align the superconducting phases with that of aluminum, whereas $J$ acts to drive the phases toward $\pi$~\cite{Supp}. This competition leads to three configurations of the relative phases (Fig. 1b-d), shown under the condition $J_1=J_2=J^{\prime}$. In the case where $\theta_1, \theta_2 \neq \phi_S$,  two possible phase angle configurations (Fig. 1d) -- related by the symmetry $\theta_i \rightarrow - \theta_j$ -- produce degenerate minima in the free energy.

The resulting Josephson effects are influenced by the competition described above. Theoretical investigations of time reversal symmetry breaking (TRSB) have been explored in junctions and interfaces between $s\pm$ and $s$-wave superconductors~\cite{Tanaka01, Ng09, Lin12, Yerin17}. The manifestation of TRSB is two-fold. First is the creation of a canted state (Fig. 1d)~\cite{Ng09, Lin12, Stanev12, Kosh12}, where a nonzero angle forms between the phase of the bands and the phase of the superconductor. The superconducting order paramter in this state is $s\pm is^{\prime}$. This canting is similar in nature to the state generated when antiferromagnetic spins are placed in a magnetic field.  The resulting effect of this canting is the generation of chiral currents in momentum space~\cite{Ng09, Lin12} that produce TRSB. Second is the generation of a predominant second harmonic in the current phase relation~\cite{Kosh12, Berg11}.  Additionally, TRSB results in four channels of supercurrent flow (Fig. 1f)~\cite{Lin12, Yerin17}: an intraband $I_{ii}$ and interband $I_{ij}$ supercurrent. The total supercurrent is governed by the phases of the two proximal conventional superconductors ($\phi_s^R-\phi_s^L$): where superscripts $R$ and $L$ denote the right and left superconducting contacts. A nonzero supercurrent will produce a relative rotation $\phi =\phi_s^R -\phi_s^L$ (Fig. 1f), thus altering each channel's relative contribution to the total supercurrent.

The phase angles in the canted state are determined by the coupling to the Al superconductor ($J^{\prime}$), which is in part determined by the density of states at the Fermi energy in the SnTe nanowire. Bulk and thin film SnTe is known to undergo a ferroelectric transition at low temperatures, causing an unequal density of states at the Fermi energy in the two bands of SnTe~\cite{Chang16, Chang19}. Hence it is important to determine whether a ferroelectric transitions occurs in SnTe nanowires. Transport measurements of the SnTe nanowires have shown clear kinks in the resistivity curves as a function of temperature, indicative of the ferroelectric transition~\cite{Shen14}. For further confirmation, the SnTe nanowires were cooled down to 12 K in \emph{in situ} cryo-transmission electron microscope (TEM) experiments to visualize the ferroelectric transition and the microstructure of the ferroelectric domains present in the SnTe nanowires at low temperature.  At room temperature, the SnTe nanowire shows uniform contrast in the bright-field TEM image (Fig. 2a); at 12 K, dark bands appear along the nanowire perpendicular to the long axis (Fig. 2b), which were absent at room temperature.  These dark bands mark the domain walls between two ferroelectric domains that emerge at low temperature.  This was confirmed by examining the electron diffraction pattern from the nanowire.  The ferroelectric transition is accompanied by the cubic-to-rhombohedral structural transition in SnTe.  As the nanowire is cooled, the cubic electron diffraction (Fig. 2c) at room temperature changes to show two sets of diffraction patterns (Fig. 2d) that are rotated by an angle of $\Delta \alpha \sim$1.2$^{\mathrm{o}}$~\cite{Supp}.  The diffraction data confirms the structural transition to the rhombohedral phase (ferroelectric phase) and suggests the presence of ferroelectric domains with primarily two domain directions, as illustrated by Fig. 2f-g.  The cubic-to-rhombohedral phase transition occurs at 80 K for this nanowire, as all the dark bands suddenly disappear at this temperature~\cite{Supp}.  We note that the dark bands are not diffraction-contrast induced contour bands as they are insensitive to swinging of the electron beam~\cite{Supp}. 

The ferroelectric distortion causes an unequal coupling ($J_1^{\prime} \neq J_2^{\prime}$) between the Al superconductor and the two bands of SnTe. The unequal coupling allows for different phase angles to form on each band ($|\theta_1| \neq |\theta_2|$)~\cite{Kosh12}. Shown in Fig. 2h-j are a comparison of the phase angles formed under equal and unequal coupling conditions. Prior to the ferroelectric transition the pocket size of the two bands is equal, producing equivalent angles with respect to the superconductor (Fig. 2h). This changes after the ferroelectric transition, where the coupling to the larger pocket is stronger, producing unequal phase angles. In general, the equilibrium phase angles $\theta_1$, $\theta_2$ will depend on the relative strength of $J, J_1^{\prime}$ and $J_2^{\prime}$.

Below we detail the manner in which the characteristic properties of SnTe JJs match the results with the above formulation. The Josephson effect of aluminum/SnTe nanowire/aluminum JJs is measured by a lock-in detection of the differential resistance $r=dV/dI$ as a function of the applied DC current ($I_{DC}$) and AC current (measured in power $P$). $r(I_{DC})$ at $P =0$ is shown in Fig. 3a. Unlike conventional overdamped JJs, different values of $I_C$ are observed for positive ($I_C^+$) and negative ($I_C^-$) $I_{DC}$. Sweeps of $I_{DC}$ in both directions reveal that the difference in $I_C^+$ and $I_C^-$ remains, confirming that the different values of $I_C$ do not arise from underdamped JJ behavior. These effects are not predicted for conventional JJs~\cite{Barone}, JJs of TCIs~\cite{Snyder18}, topological insulators~\cite{Williams12, Veldhorst12, Hart14}, or strong spin-orbit nanowires~\cite{Zuo17}. A current-direction-dependent $I_C$ has also been observed in junctions where time reversal symmetry is broken: this has been observed junctions containing a ferromagnetic weak link~\cite{Gold07, Gold11, Sickinger12}. In addition to unequal critical currents, we observe an anomalous magnetic diffraction pattern in the DC Josephson, which indicates the presence of two channels of supercurrent that are $\pi$ out of phase with each other~\cite{Supp}. 

\begin{figure}[t!]
\center \label{fig3}
\includegraphics[width=3 in]{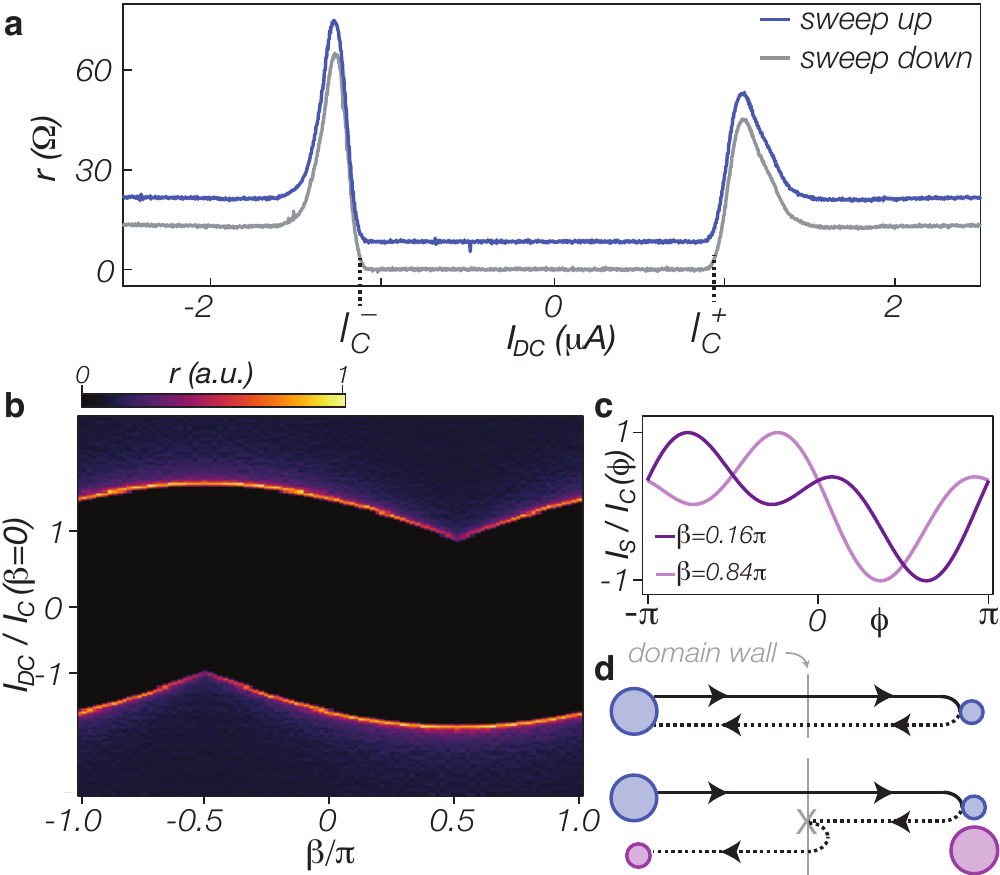}
\caption{\footnotesize{\textbf{Breakdown of the DC Josephson Effect.} \textbf{a}, Differential resistance $r$ as a function of DC bias current $I_{DC}$ in different sweep directions. The bias sweeps show no hysteresis and two nonidentical critical currents $I_C^+$ and $I_C^-$. The curves are offset for clarity. \textbf{b}, The simulated differential resistance $r(I_{DC},\beta)$ calculated by the resistively-shunted junction model using a CPR of $I_S=\sin(\phi)+A\sin(2\phi+\beta)$, where the best fit parameters $\beta$ with the experiment are $0.16\pi$ and $0.84\pi$. The resulting CPRs are plotted in \textbf{c}.  \textbf{d}, Demonstration of inequivalent phase accumulation for electron (solid line) and hole (dashed line) Andreev bound state pairs. Pairs which remain in the same band (bands are shown by blue and purple dots) do not acquire an additional phase from the domain wall (upper), whereas scattering of either the electron or hole trajectory at the domain wall (lower) incurs an additional phase. }}
\end{figure}

To understand the origin of the difference between $I_C^+$ and $I_C^-$, numerical simulations of the resistively-shunted junction model~\cite{Supp} were performed (Fig. 3b).  Conventional JJs possess a current-phase relation (CPR) $I_S(\phi)$, which is both inversion and $\pi$-translation symmetric, a result of time-reversal symmetry. The only way to reproduce $r(I_{DC})$ curves that are not symmetric in $I_{DC}$ is to break both of these symmetries: this simplest CPR that accomplishes this is $I_S=\sin(\phi)+A\sin(2\phi+\beta)$, where $\beta$ is a fit parameter and $A=0.909$ is determined by the AC Josephson effect [the second harmonic is expected to be predominant in the TRSB state~\cite{Kosh12, Sperstad09} and will be confirmed in our measurement of the AC Josephson effect (Fig. 4)]. Two values of $\beta=(0.16,0.84)\pi$ best match the experimental data.  The CPRs for these values of $\beta$ are shown in Fig. 3c: importantly, these CPRs break time-reversal symmetry, i.e. $I(\phi) \neq - I(-\phi)$. The essential features of these CPRs are the two minima/maxima which occur at different values of $I_S$: it is these features which give rise to the differences in $I_C^+$ and $I_C^-$.  CPRs with similar minima/maxima characteristics have been predicted in JJs that share some of the characteristics of the JJs consider in this work. These are multiband JJs with unequal coupling~\cite{Sperstad09}, Josephson junctions between an s-wave and a three band superconductor~\cite{Huang14}, and in JJs across domain walls in unconventional superconductors~\cite{Tanaka97}. 

The coexistence of 4 supercurrent channels (Fig. 1f) is expected to result in a reduction of the critical current since 0 and $\pi$ channels spatially coexist~\cite{Lin12, Linder09} and the supercurrent is carried entirely by the second harmonic term~\cite{Kosh12, Berg11, Sperstad09}.  The ferroelectric domain walls thus serve two critical functions in the modification of supercurrent through the JJ. The first is to enhance the relative number of carriers that cross the domain wall while scattering between bands~\cite{Chang19}. This is necessary to establish negative interband coupling $J$. The second is the domain wall allows for supercurrents that have a phase offset with respect to each other (needed to create the unequal minima/maxima that breaks $\pi$-translation symmetry).  In Fig. 3d, we show how this occurs. The condition for bound states in JJs arises from the criterion that the round-trip phase accumulation $\chi_{tot}$ of the electron/hole pair is 2$\pi$: for conventional junctions in the short junction limit ($L \ll \xi_o$, where $\xi_o$ is the coherence length of the leads and $L$ is the length of the JJ) this criterion is given by: $\chi_{tot}=2\text{arccos}(E/\Delta) + \phi$, where $\Delta$ is the size of the induced superconducting gap and $E$ is the energy of the electron/hole relative to the Fermi energy. CPRs derived from this have zero offset phase. Crossing the domain introduces additional phases to this equation. First consider the case where the electron/hole remain in band 1 (upper diagram in Fig. 3d). Crossing the domain wall induces a phase shift because the phases in band 1 are unequal on either side of the domain wall (Fig. 2i,j). If the hole remains in the same band, the phase accumulated across the domain wall is the opposite. Thus, the additional phases arising from the domain wall cancel and a CPR zero offset phase results. If, however, a single scattering event occurs at the domain wall (lower diagram in Fig 3d), the hole accumulates a difference phase than the incident electron and the criterion for a bound state becomes:  $\chi_{tot}=2\text{arccos}(E/\Delta) + \phi +\chi^{DW}$, where $\chi^{DW}$ is the difference in phase accumulation~\cite{Supp}. These trajectories produce a CPR with a phase offset of $\chi^{DW}$. 

We now turn our attention to the modification of the AC Josephson effect. The presence of a second harmonic component -- expected in the TRSB state~\cite{Kosh12, Berg11, Sperstad09} --  will result in additional steps at values of half the expected $hf/2e$. A plot of $r(I_{DC}, P)$ is shown in Fig. 4a taken at $f$=5\,GHz. In addition to dips in $r$ observed at the expected integer values (labeled in white), prominent features at half-integer values are also apparent.  This is more clearly seen in cuts of Fig. 4a, shown in Fig. 4b taken at $P$=-11.8dBm. In addition to the dips in $r$ (grey curve) at integer values, clear dips at half integer values occur. In fact, the drop in $r$ at $1/2$ is nearly equal to that at $1$. In addition, the integrated voltage $V = \int (\sfrac{dV}{dI})dI$ versus $I_{DC}$ curve is shown in blue. The dips/plateaus measured  in $r$/$V$ are nearly equal in strength, indicating that the contributions of the first and second harmonic to the CPR are approximately equal: the depth magnitude is used to extract the value of $A=0.909$ for the CPR used in the numerical simulations of Fig. 3b. 

The Shapiro diagram (Fig. 4a) also has two other signatures that indicate nearly equal contribution from a first and second harmonic term. First, the width of the zeroth step does not go to zero (indicated by the two white vertical lines), as expected for the zeroth order Bessel function. It does go to zero for the second closure. Second, hile the step width in $I_{DC}$ of the half integer steps is modulated with $P$, showing regions of $P$ where the step width goes to zero (as is expected), the width modulation is less pronounced on the integer steps. These differences occur when the CPR has both first and second harmonic terms~\cite{Supp}.

\begin{figure}[t!]
\center \label{fig4}
\includegraphics[width=3 in]{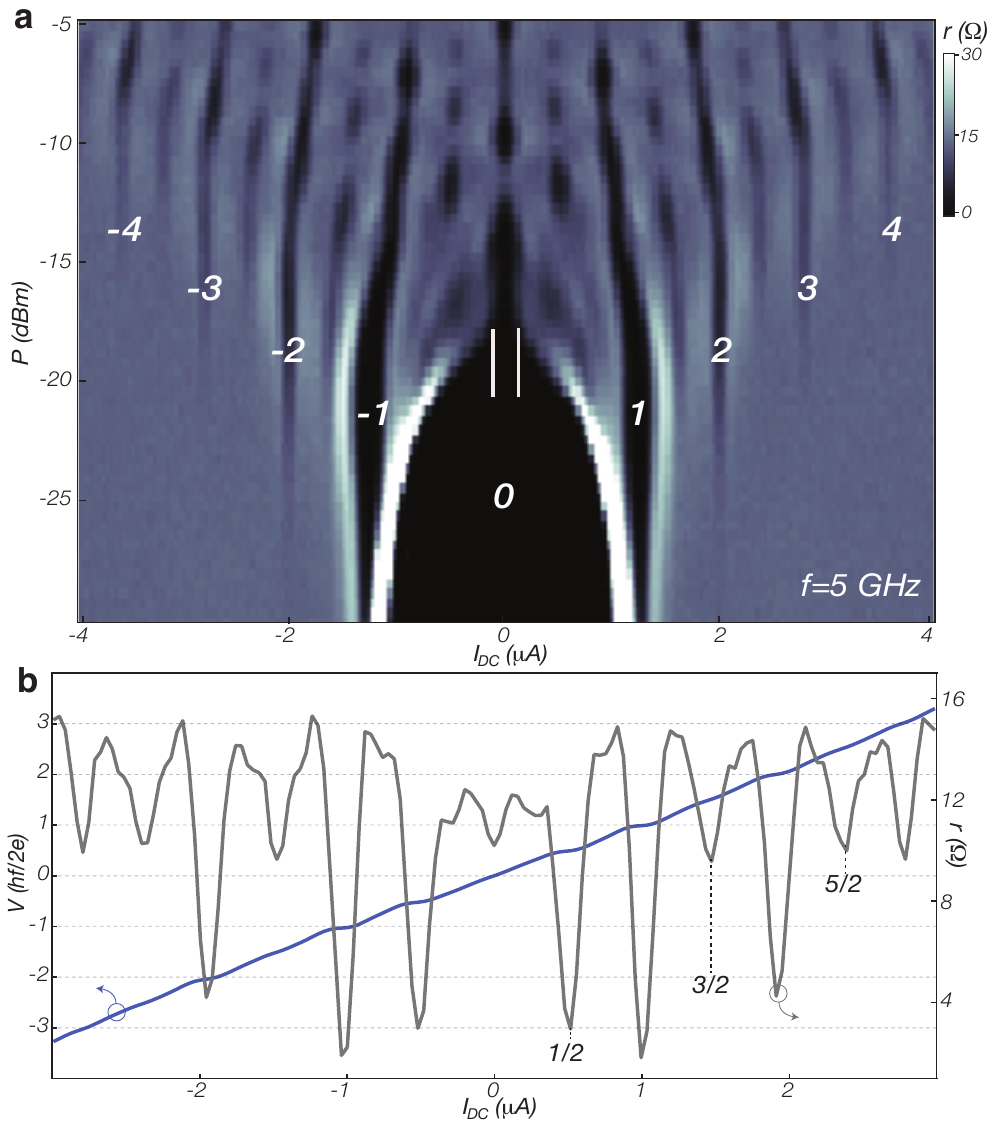}
\caption{\footnotesize{\textbf{AC Josephson Effect.} \textbf{a}, In addition to the integer Shapiro steps, labeled in white, fractional Shapiro steps appear in between the integer ones in a plot of $r(\Idc, P)$. \textbf{b}, $r(I_{DC})$ (grey curve) and integrated voltage $V(\Idc)$ (blue curve) taken at an applied RF frequency $f$ = 5 GHz, $P$ = -11.8 dBm. The first 1/2 integer step occurs with a nearly equal intensity to the first integer step. }}
\end{figure}

Subharmonic steps are expected for underdamped junctions and for overdamped junctions with a skewed CPR. Our junctions are overdamped; hence we rule out  the former as being the cause. Skewed CPRs in overdamped junctions produce fractional Shapiro steps, but the strength of these steps is much reduced compared to the integer steps. For comparison, see Ref.~\cite{Snyder18} for the AC Josephson effect in a similar material,  Pb$_{0.5}$Sn$_{0.5}$Te). The CPR in Ref.~\cite{Snyder18} used to reproduce the experimental data is that of a ballistic JJ, a CPR which has a greatest amount of skewness amongst the candidate CPRs. Yet, it produces dips at fractional values that are an order of magnitude smaller than the integer value dips.  Therefore, we also rule out the skewed CPR as the source of the observed effect.

In summary, we have investigated the combined effects of proximity-induced multiband superconductivity and ferroelectric distortion on the dynamic properties JJs of SnTe nanowire weak links. Such effects offer new routes to control the flow of supercurrents, where modification of the density of states or ferroelectric transition temperature by electric fields and strain can be used to modulate the supercurrent and the offset phase in the device. The manifestation of multiband and multicomponent superconductivity in our devices offers experimental access to the phase induced on individual bands. This allows for the investigation of the order parameter in novel superconductors~\cite{Tanaka15, Yerin17}, like iron-based superconductors, and for the determination of topology in the superconducting state~\cite{Qi10, Hosur14, Guguchia17}. \\

\noindent \emph{Materials and Methods}

SnTe nanowires measured in the study were synthesized by metal-catalyzed chemical vapor deposition using a single-zone furnace.  SiO$_2$/Si substrates decorated with 20 nm-wide gold nanoparticles were used as growth substrates.  SnTe and Sn source powders were mixed and placed at the center of a horizontal quartz tube with 1-inch diameter while the growth substrates were placed upstream in the quartz tube, 10-13 cm away from the center.  The furnace was heated to 600$^o$C and remained at the temperature for 1hr with an Ar carrier gas at a flow rate of 20 s.c.c.m.  After the growth, the furnace was allowed to cool naturally.  The growth substrates contain SnTe microcrystals, nanoplates, and nanowires whose atomic structure and chemical composition were characterized by transmission electron microscopy and energy dispersive X-ray spectroscopy~\cite{Supp}.  For Josephson junction studies, we select SnTe nanowires with diameters  $< \sim$300 nm.  The details of the synthesis reactions and microcharacterizations of SnTe nanowires can be found in our previous reports (Refs. [S10]) and Ref.~\cite{Supp}.  

The \emph{in-situ} cryo-TEM experiments were carried out using Gatan's liquid-He cryo holder (HCTDT 3010) and JEOL JEM-ARM200CF at 200 kV at Brookhaven National Laboratory.  SnTe nanowires were drop-casted onto Cu-mesh TEM grids overlaid with a thin carbon support film.  The TEM sample was cooled from room temperature to 12 K by cooling the cryo holder with liquid helium.  The temperature sensor measures the temperature of the holder, and the actual temperature of the sample may be ~$\sim$ 5--10 K higher.  The \emph{in-situ} cryo TEM movie was acquired by naturally warming the TEM sample.  During cryo-TEM experiments, electron-beam damage was observed in SnTe nanowires when they were exposed to the electron beam for a prolonged time~\cite{Supp}. 

Transport measurements were carried out in a dilution refrigerator with a base temperature of 25mK. DC electrical leads were heavily filtered to remove high frequency noise above 10kHz. Lock-in detection of the differential resistance was carried out using a 1nA excitation at 13Hz. Radio frequency radiation up to 7GHz was supplied to one of the electrical leads via a synthesizer through a bias-tee located on the chip carrier.\\

\emph{Acknowledgments}:  Synthesis of narrow SnTe nanowires was supported by NSF 1743896.  Transport characterizations of SnTe nanowires were supported by DOE DE-SC0014476. The cryo-TEM work was supported by the US DOE Basic Energy Sciences, Materials Sciences and Engineering Division under Contract No. DE-SC0012704. Transport measurements of the JJ devices were sponsored by the grants National Science Foundation A ``Quantum Leap" Demonstration of Topological Quantum Computing (DMR-1743913),  Physics Frontier Center at the Joint Quantum Institute (PHY-1430094), and the Army Research Office Award W911NF-18-2-0075.

\onecolumngrid
\appendix
\newpage 

\noindent \large \textbf{Supplementary Information for Josephson Detection of Time Reversal Symmetry Broken Superconductivity in SnTe Nanowires} 

\normalsize
{
\section{Model Hamiltonian and Mean-Field Solutions}}
We consider the following model Hamiltonian with single-electron part $H_0$ and interactions
\begin{eqnarray}\label{eq_H}
H=H_0+\frac{1}{2}\sum_{j=1,2}(h_{j} c_{j\uparrow}c_{j\downarrow}+ Uc_{j\downarrow}^{\dagger}c_{j\uparrow}^{\dagger}c_{j\uparrow}c_{j\downarrow}+h.c.)+\frac{1}{2}(gc_{1\downarrow}^{\dagger}c_{1\uparrow}^{\dagger}c_{2\uparrow}c_{2\downarrow}+h.c.)
\end{eqnarray}
where $j=1,2$ is the pocket index, $h_{j}$ is the induced pairing in pocket $j$, $U$ is the intrapocket density-density interaction and $g$ is the interpocket interaction. Without external superconductors, there is no intrinsic pairing and $U>|g|>0$. With external superconductors, Coulomb interaction can be screened and in the following we assume $ 0<U<|g| $. 

We would like to use the following mean-field Hamiltonian to approximate the model Hamiltonian
\begin{eqnarray}\label{eq_MFh}
H_{\rm MF}=H_0+\frac{1}{2}\sum_{j=1,2}(\Delta_j c_{j\uparrow}c_{j\downarrow}+h.c.)
\end{eqnarray}
where all interactions and induced pairing contribute to intrapocket pairing potentials $\Delta_j$.

%
%

By taking the mean-field average of the model Hamiltonian
\begin{eqnarray}\label{eq_HMF}
H_{\rm MF}=\langle H\rangle =H_0+\frac{1}{2}\sum_{j=1,2}(\Delta_0 c_{j\uparrow}c_{j\downarrow}+ U\langle c_{j\downarrow}^{\dagger}c_{j\uparrow}^{\dagger}\rangle c_{j\uparrow}c_{j\downarrow}+g\langle c_{\overline{j}\downarrow}^{\dagger}c_{\overline{j}\uparrow}^{\dagger}\rangle c_{j\uparrow}c_{j\downarrow}+h.c.)
\end{eqnarray}
we obtain the mean field equation of model Hamiltonian (\ref{eq_H})
\begin{eqnarray}\label{eq_MF}
\bm{\Delta} =\hat{I}\bm{\Psi} +\bm{h}
\end{eqnarray}
where the pairing potential vector $ \bm\Delta $, pairing correlation vector $\bm\Psi$, external pairing field $\bm h$ and interaction matrix $\hat{I}$ are
\begin{eqnarray}
\bm{\Delta} =
\begin{pmatrix}
\Delta_{1}\\
\Delta_{2}
\end{pmatrix},\quad
\bm{\Psi} =
\begin{pmatrix}
\Psi_1 \\
\Psi_2
\end{pmatrix}
,\quad
\bm{h}=
\begin{pmatrix}
h_1\\
h_2
\end{pmatrix},\quad
\hat{I}=
\begin{pmatrix}
U&g\\
g&U
\end{pmatrix},
\end{eqnarray}
and the pairing correlation of pocket $j$ reads
\begin{eqnarray}\label{eq_pc}
\Psi_{j}\equiv\langle c_{j\downarrow}^{\dagger}c_{j\uparrow}^{\dagger}\rangle =-\int_{-\infty}^{\infty} \frac{\rho\Delta_j}{\sqrt{\xi^2 +|\Delta_j |^2}}\tanh\frac{\sqrt{\xi^2 +|\Delta_j |^2}}{2T}d\xi .
\end{eqnarray}

\subsection{Numerical Results}
The mean-field equation (\ref{eq_MF}) can be regarded as a fixed point equation of the mapping $ M:\bm\Delta\to\hat{I}\bm\Psi(\bm\Delta) +\bm h $, which can be solved by numerical iteration
\begin{eqnarray}
\bm\Delta =\lim_{n\to\infty}M^n(\bm\Delta_0)
\end{eqnarray}
with appropriate initial guess $\bm\Delta_0$. To make the iteration unbiased, we choose the complex initial guess  $\bm\Delta_0\in\mathbb{C}^2$. We write the solutions as the amplitude and the phase $\Delta_j=|\Delta_{j}|\exp(i\theta_j)$ in pocket $j=1,2$.

From the numerical results, we find for some external pairing fields $\bm h$, the solutions to the mean-field equation (\ref{eq_MF}) can be complex, and the phase difference between the two pockets varies from 0 to $\pi$, as shown in Fig. S1. This can be explained by the Josephson part of the free energy as shown below. 

\renewcommand{\thefigure}{S1}
\begin{figure}
\includegraphics[width=0.75\textwidth]{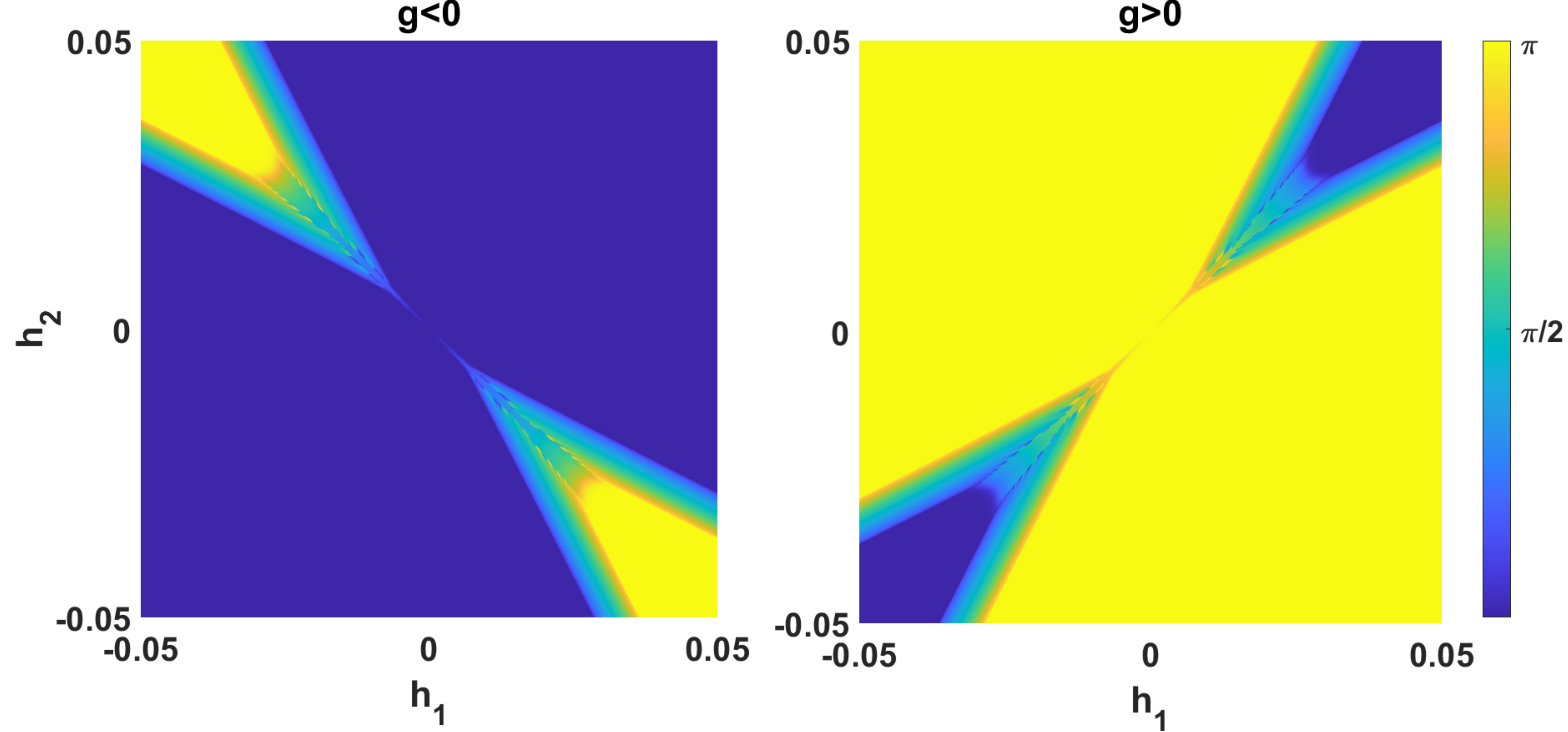}
\centering
\caption{{Phase difference $\theta_1-\theta_2$ of the ground state solution to mean-field equation Eq. (\ref{eq_MF}). Here $ U=0.1,|g|=0.12,T=10^{-4} $, and the numerical integration interval of pairing correlation (\ref{eq_pc}) is from $ -100 $ to $ 100 $.}}\label{figS9}
\end{figure}

\subsection{Phenomenological Theory}
We consider Ginzburg-Landau free energy of the following form
\begin{eqnarray}\label{eq_f0}
F=(\bm d\cdot\bm\Delta +c.c.)+\frac{1}{2}\bm\Delta^{\dagger}A\bm\Delta +\sum_{i=1,2}b_{i}|\Delta_{i}|^4,\quad A=
\begin{pmatrix}
a_1 & c\\ 
c & a_2
\end{pmatrix},
\end{eqnarray}
where $\bm d\in\mathbb{R}^2$ is due to external pairing fields, $A$ is the Hessian matrix and $ b_{i}>0 $ to stablize the free energy. In fact we can always rescale the order parameters $ \Delta_{i}\to\sqrt[4]{b/b_{i}}\Delta_{i} $ to make the quartic coefficients the same $ b_{1}=b_{2}\equiv b $.

In terms of phase and amplitude $ \Delta_{j}=|\Delta_{j}|e^{i\theta_{j}} $, the free energy can be rewritten as the Josephson and amplitude parts
\begin{eqnarray}\label{eq_f1}
F&=&F_{J}+F_{0},\\
F_{J}&=&J\cos(\theta_1-\theta_2)+J_{1}\cos{\theta}_1+J_{2}\cos\theta_{2},\\
F_{0}&=&\sum_{i=1,2}\left\{\frac{1}{2}a_{i}|\Delta_{i}|^2+b_{i}|\Delta_{i}|^4\right\},
\end{eqnarray}
where
\begin{eqnarray}
J=c|\Delta_{1}\Delta_{2}|,\quad J_{i}=d_i|\Delta_i|.
\end{eqnarray}
We first minimize the Josephson energy to obtain $\theta_{1,2}=\theta_{1,2}^{*}$ under the assumptions $ J,J_{1,2}\neq 0 $, then we minimize $ F_{0}+F_{J}^{*} $ to obtain $|\Delta_{j}|=\Delta_{1,2}^{*}$, where $ F_{J}^{*}=F_{J}|_{\theta=\theta^{*}} $.
When $ \Delta_j^{*}\neq 0 $, the solutions of $\theta_{1,2}^{*}$ are consistent with $ \Delta_j^{*} $; otherwise they are rejected due to self inconsistency.

By minimizing the Josephson energy alone, we find the phase difference $ \theta\equiv\theta_1-\theta_2 $ bewteen two pockets is ($ \Theta $ is the step function with $ \Theta(0)=1 $)
\begin{eqnarray}\label{eq_theta}
\theta ={\rm Re}\left(\arccos{\left[\frac{J_{1}^2J_{2}^2-(J_{1}^2+J_{2}^2)J^2}{2J_{1}J_{2}J^2}\right]}\right)\Theta(JJ_{1}J_{2})+\pi\Theta(J)\Theta(-J_1J_2).
\end{eqnarray}

When $ \theta\neq 0,\pi $, the minimal Josephson energy reads
\begin{eqnarray}
F_{J}^{*}=-\frac{1}{2}c\left(\frac{d_1d_2}{c^2}+\frac{d_1}{d_2}|\Delta_{1}|^2+\frac{d_2}{d_1}|\Delta_{2}|^2\right),
\end{eqnarray}
and hence
\begin{eqnarray}
F_{0}+F_{J}^{*}=\sum_{i=1,2}\left\{\frac{1}{2}a_{i}^{*}|\Delta_{i}|^2+b_{i}|\Delta_{i}|^4\right\},\quad a_{1}^*=a_{1}-c\frac{d_1}{d_2},\quad a_{2}^*=a_{2}-c\frac{d_2}{d_1}.
\end{eqnarray}

The self-consistency condition requires that
\begin{eqnarray}
a_{1,2}^*<0.
\end{eqnarray}
When $ a_{1,2}>0 $ and $c>0$ this leads to
\begin{eqnarray}
\frac{a_1}{c}<\frac{d_1}{d_2}<\frac{c}{a_2}\Rightarrow a_1 a_2<c^2,
\end{eqnarray}
and when $ a_{1,2}>0 $ and $c<0$ this leads to
\begin{eqnarray}
\frac{a_1}{c}>\frac{d_1}{d_2}>\frac{c}{a_2}\Rightarrow a_1 a_2<c^2.
\end{eqnarray}
In these two cases, $ A $ has one negative eigenvalue. This corresponds to the assumption $0<U<|g|$ we imposed in the beginning, and explains the finite phase difference between pairing order parameters in two pockets.

In the case where the coupling to each band is equal ($J_1=J_2=J^{\prime}$), the dependence of the angle between the bands on the ratio $J^{\prime}/J$ is
\begin{equation}
\theta = 2\text{Re}(\text{arccos}(\delta/2)), \hspace{0.25cm} \delta=\lvert \frac{J^{\prime}}{J} \rvert.
\end{equation}  

The dependence of $\theta$ on $J^{\prime}/J$ is shown in Fig. 1e of the main text.

\renewcommand{\thefigure}{S2}
\begin{figure}
\includegraphics[width=0.75\textwidth]{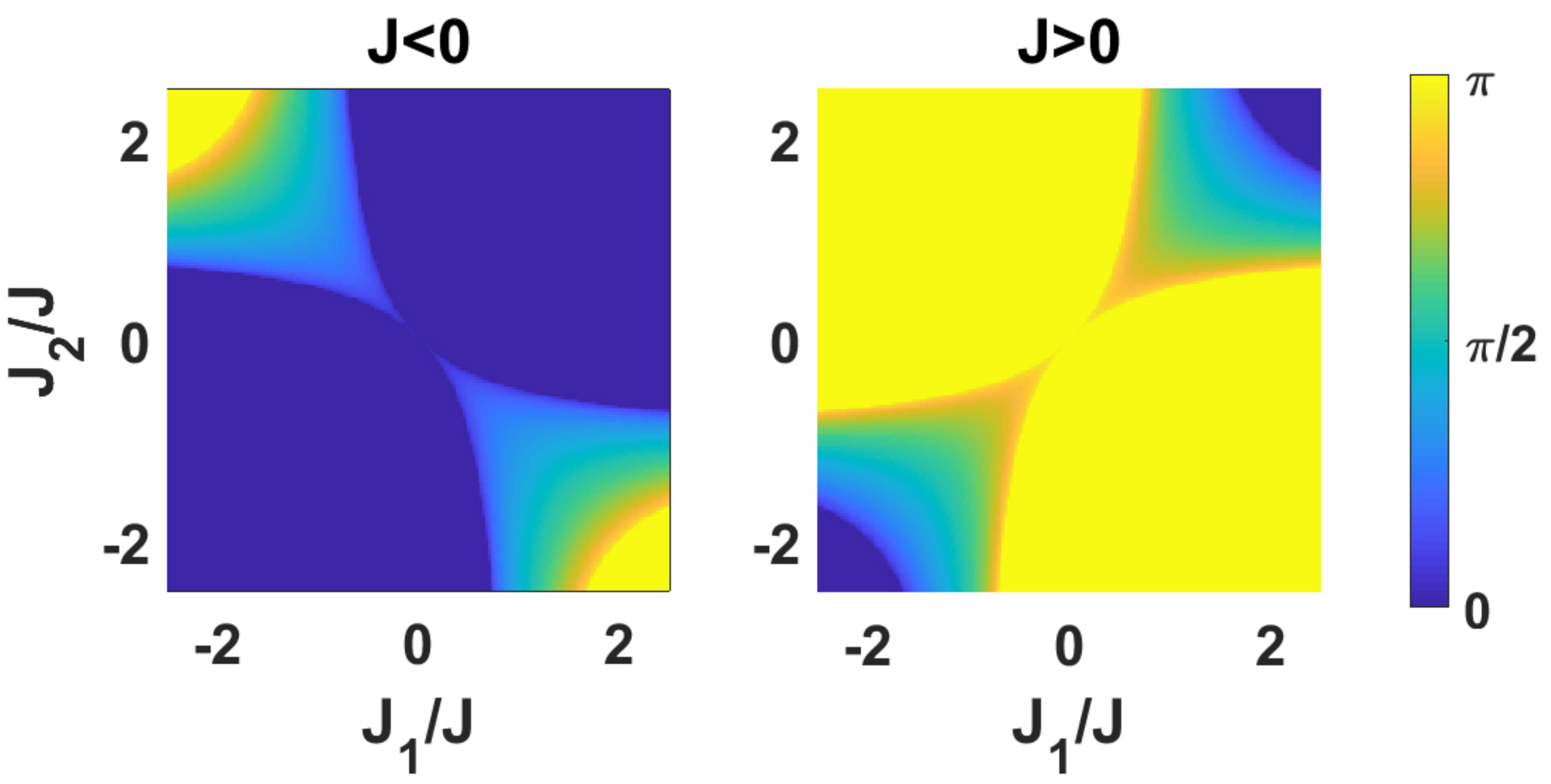}
\centering
\caption{{Phase difference $\theta_1-\theta_2$ of the ground state of Josephson free energy $F_J$ in Eq. (\ref{eq_f1}).}}\label{figS10}
\end{figure}

\section{Calculation of the Phase Shift of the Josephson Currents}
In this section we examine the effect of the phase shift at a domain wall on the Andreev bound states states and the zero temperature current phase relation (CPR). It is first important to recall the argument used to arrive at the CPR for a ballistic conductor. The condition for constructive addition of the electron/hole pair wavefunction in the junction occurs when the total round-trip phase accumulated is $2\pi$: it is these resonant states that give rise to Andreev bound states. The phase acquire in an Andreev reflection of an electron into a hole ($\chi_{eh}$) and a hole into an electron ($\chi_{he}$) is given by:

\renewcommand{\thefigure}{S3}
\begin{figure*}[t]
\center \label{fig:S3}
\includegraphics[width=0.75\linewidth]{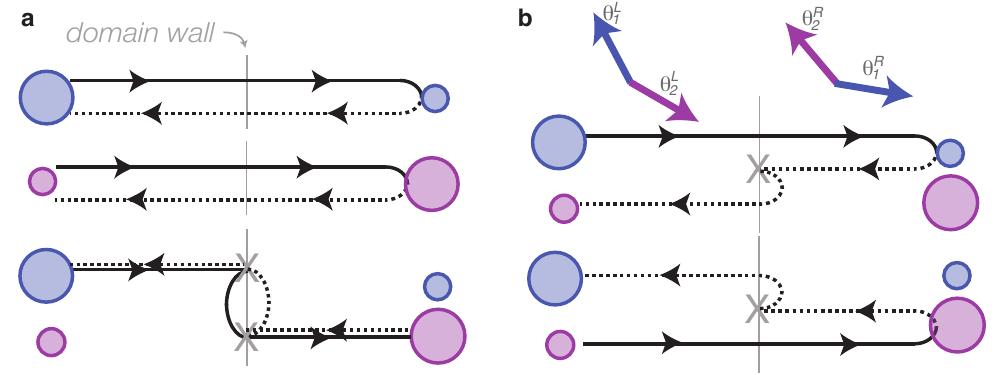}
\caption{\footnotesize{A comparison of electron/hole trajectories across the domain wall. Bands 1(2) are indicated blue(purple) and the size of the circle indicates the density of states at the Fermi level. Electron (hole) trajectories are shown in solid (dashed) lines and arrows indicate direction of propagation. \textbf{a}, For trajectories which involve no interband scattering at the domain wall or an even number of scattering events, no net accumulation of phase at the domain wall occurs. Hence, supercurrent generated by these trajectories produce a CPR with no phase offset. \textbf{b}, for an odd number of interband scatterings, a net phase is accumulated for transport across the domain wall. Upper diagram shows the phase angles on either side, expected from Fig. 2 of the manuscript. }}
\end{figure*}

\begin{equation}
\chi_{eh} = -\text{arccos}(E/\Delta)+\phi_s, \hspace{0.2cm} \chi_{he} = -\text{arccos}(E/\Delta)-\phi_s, 
\end{equation}

\noindent where $E$ is the energy of the electron/hole state in the weak link, $\Delta$ is the magnitude of the superconducting gap and $\phi_s$ is the phase of the superconductor that produced the reflection. Setting the total accumulated phase $\chi_{tot}=\chi_{eh} +\chi_{he}$ equal to $2\pi$ produces Andreev bound states of a ballistic junction $E(\phi)=\pm \Delta \text{cos}(\phi/2)$. The CPR at zero temperature is determined by $I_S(\phi)=dE/d\phi$.

The presence of a domain wall allows for scattering between bands. As the electron/hole traverses the domain wall, an addition phase will be acquire if the bands involve in transporting the electron/hole have different phases. First consider that the electron/hole remain in the same band when crossing the domain wall (upper portion of Fig. S3a). Following the trajectory indicated by the arrows, an electron starts in band 1 on the left side of the wall, crosses the wall and returns to band 1 as a hole. Since band 1 has different phases across the domain wall (see upper portion of Fig. S3b), the electron acquires an addition phase $\theta_1^R-\theta_1^L$. The returning hole acquires the opposite phase $\theta_1^L-\theta_1^R$, and the net round trip phase acquired at the domain wall is zero. The same occurs for two interband scattering events shown in the lower portion of Fig. S3a. Hence the CPR determine here will have zero phase offset. 

The situation changes where there is an odd number of interband scattering events at the domain wall during the round trip of the electron hole pair. Shown in Fig. S3b are 2 of the 8 possible electron/hole trajectories with an odd number of scatters at the domain wall. Starting with the scattering trajectories of the middle diagram of Fig. S3b -- i.e. a path which starts as an electron in band 1 on the left and returns to band 2 on the left -- the phase accumulated at the domain wall is $\chi^{DW}=\theta_1^R-\theta_1^L$ for the electron and $\theta_2^L-\theta_1^R$ for the hole. These phases do not cancel and the total accumulated phase and Andreev bound state spectrum are: 

\begin{equation}
\chi_{tot}= -2\text{arccos}(E/\Delta)+\phi-\theta_2^L-\theta_1^L, \hspace{0.2cm} E(\phi)=\pm \Delta \text{cos}[(\phi + \theta_2^L-\theta_1^L)/2].
\end{equation}

\noindent Hence, the Andreev bound state for this trajectory acquires a phase offset of $\theta_2^L-\theta_1^L$, related to the difference of the phases of the two bands on the left side of the domain wall. For each of the 8 trajectories, there are complimentary trajectories that accumulate the exact opposite phase. The trajectory complimentary to the one just described is shown in the lower portion of Fig. S3b. Here $\chi^{DW}$ is $\theta_1^L-\theta_2^L$. The current arising from these two process do not cancel however. The reason can be seen by looking at the relative weight of each excursion. The pair of trajectories producing the phase shift of Eq. 22 originate from the larger of the two bands, which possess a large density of states at the Fermi energy. the supercurrent transmitted from this Andreev bound state is larger, since more carriers originate from band 1. Hence, when summed together, supercurrent from these two complimentary trajectories produces an overall nonzero phase shift in the CPR. 

\section{Device Fabrication and DC Josephson Charateristics}

\renewcommand{\thefigure}{S4}
\begin{figure}[h]
\center \label{figS4}
\includegraphics[width=0.5\linewidth]{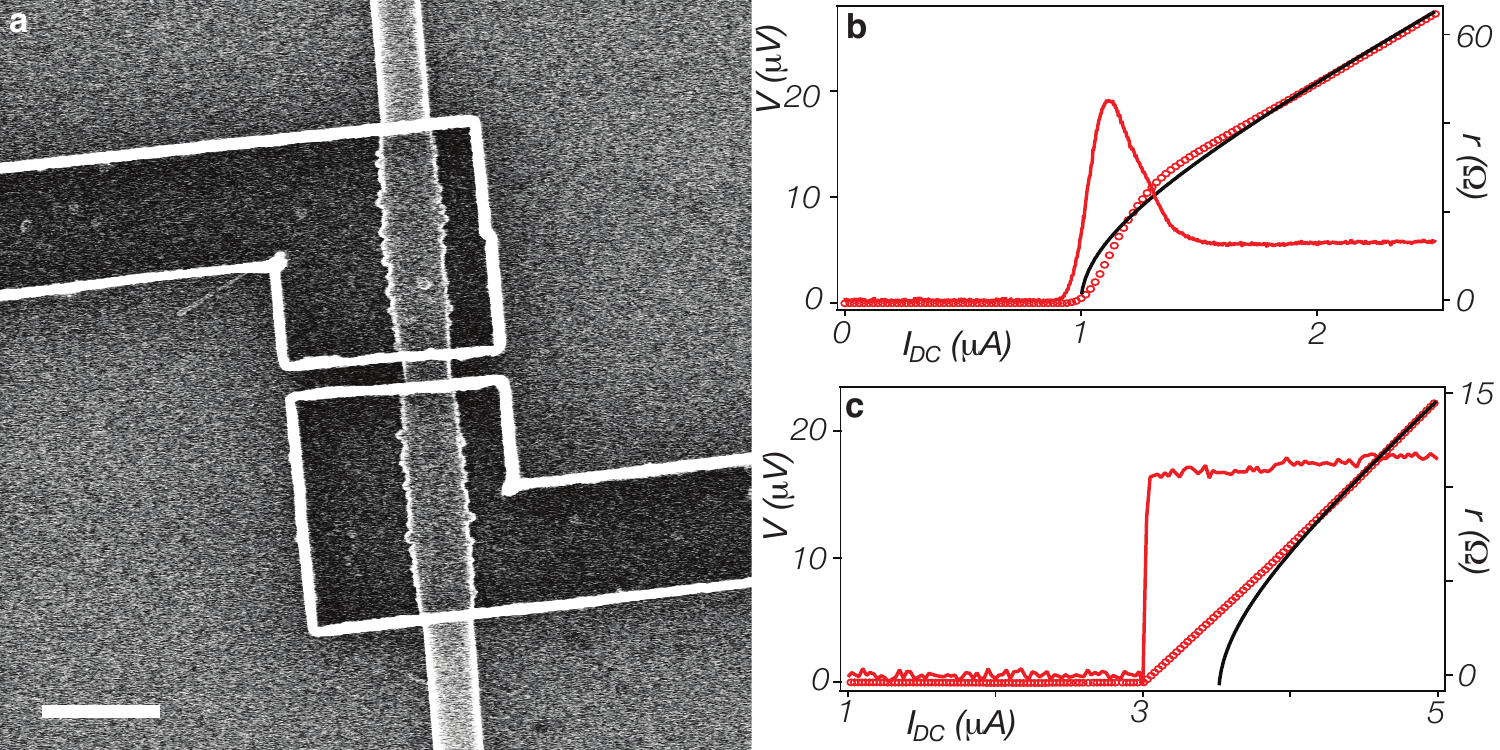}
\caption{\footnotesize{\textbf{a}, A scanning electron micrograph of a Josephson junction atop a SnTe nanowire. The scale bar shown in white is 1 $\mu$m. \textbf{b, c}, A comparison of overdamped junction behavior (\textbf{b}) and switching junction behavior seen in the measurement of $r$. In \textbf{b, c} the red line is $r$, red circles is $V$ obtained from integration of $r$ and the black line is the expected voltage from the solution of the RSJ model. Good agreement between the RSJ model and the junction in \textbf{b} is observed.}}
\end{figure}

Our Josephson devices are fabricated on $\sim$ 5mm x 5mm Si/SiO$_2$ chips with SnTe nanowires dispersed atop them. First, a pattern of equally spaced alignment marks is written using electron-beam lithography with a dose of 1600 $\mu$C/cm$^2$. After a 60s {\it{in situ}} argon plasma etch at 50W, Ti/Au are deposited (5 nm/70 nm) using e-beam evaporation. 

After liftoff of the alignment marks, ideal SnTe wires are selected using an optical microscope. Then, the Josephson devices are written atop these wires using a dose of 1600 $\mu$C/cm$^2$. The sample undergoes a 60s {\it{in situ}} argon plasma etch at 50W, followed by the sputtering of Ti/Al (4.5 nm/200 nm). An essential part of getting samples with measurable supercurrents at base temperature is the heating of the sample during deposition of aluminum. During the deposition of aluminum, the sample is heated to 100\degree C. A scanning electron microscope image of a completed device is shown in Fig. S4.

Important in the discussion of critical currents is the nature of the transition between the super and normal current flow. Already demonstrated in the main text is the lack of hysteresis, shown in the identical plots of $r(I_{DC})$ for each current ramp direction; the lack of hysteresis means the junction is overdamped (Ref. [S1]). The relationship between voltage $V$ and current $I_{DC}$ in overdamped junctions can be obtained from the solution of RSJ model (Eq. 23, below): $V(I)=R(I_{DC}^2-I_C^2)^{1/2}$. In Fig. S4b we compare $V(I)$ for the sample in the main text with the RSJ model. The results show good agreement, indicating that abrupt switching does not play a significant role in the transition. This can be contrast with junctions of materials where switching plays a key role. Shown in Fig S4c is an Al-WTe$_2$-Al junction measured in the same setup as the device in the main paper. Rather than observing a gradual transition in $r$ at $I_C$, an abrupt jump in $r$ is measured. This jump is associated with a slip in the phase, cause either by thermal fluctuations of the phase or macroscopic quantum tunneling. This jump in $r$ is not capture by the RSJ model, as seen in Fig. S4c. 

Comparisons of this type are important, as the presences of competing ``0" and ``$\pi$" channels in the flow of supercurrent invalidates the use of $I_C$ as a metric of the induced superconducting gap and the Josephson energy: in conventional junction,  the induced gap is given by $I_CR_N \propto \Delta/e$ and the Josephson energy is $E_J=\hbar I_C/2e$. As shown in Ref. 30 of the main text, the prominences of a second harmonic stems from a suppression of the first harmonic from the competing ``0" and ``$\pi$" channels. This suppression will reduce the value of $I_C$ but not the values of the induced gap. A similar phenomena occurs in JJs with ferromagnetic weak links. 

\section{More Data on Shapiro Steps}

The Shapiro step pattern observed in our devices is influenced by the applied perpendicular magnetic field. Fig. S5 shows maps of differential resistance measured at 2.5GHz as a function of RF power and $I_{DC}$ at magnetic fields of 0 to 46mT plotted in the same color scale. These RF maps reveal several features. First is the subharmonic steps seen at $B$ = 0 deepen as the field approaches 16 mT, the field of maximum $\Ic$. At higher fields they diminish again before disappearing fully. Second, the minimum RF power of the closed $N=0$ Shapiro step also changes with the magnetic fields in the same manner as the magnetic diffraction pattern (see below). This means the critical currents can also define the maximum amplitude of the driven RF current to stay in the superconducting state. Third, the first steps $N=\pm1$ merge with the second steps when the magnetic field is between 8mT to 20mT. Last, at some magnetic fields we also observe the presence of an arc breaking through and disturbing the pattern at lower powers. This ``broken ribs" feature is also most prominent at 16 mT and only observed at an RF frequency of 2.5 GHz, not at other frequencies we measured (see Fig. S6). The origin of the effect remains an open question.

Fig. S6 shows additional RF maps measured at different frequencies and fields. Fig. S6 a-c and e are taken at 16mT and frequencies of 2, 3, 4 and 5GHz, respectively. Generally, half steps at higher frequencies are deeper and wider. Along with the 2.5GHz, 16mT map in Fig. S5, these maps clearly show that the merging of the first and second Shapiro steps are only present at low frequencies (2GHz and 2.5GHz), consistent to a recent experiment (Refs. [S3-S5]). Fig. S6d,e compare the RF maps taken at 5GHz at 0mT and 16mT. The critical currents at 16mT are also larger than those at 0mT. Moreover, fainter 1/3 steps appear near -17dBm at 0mT (Fig. S6d), but get suppressed at 16mT. 

\renewcommand{\thefigure}{S5}
\begin{figure*}[t]
\center \label{figS5}
\includegraphics[width=\linewidth]{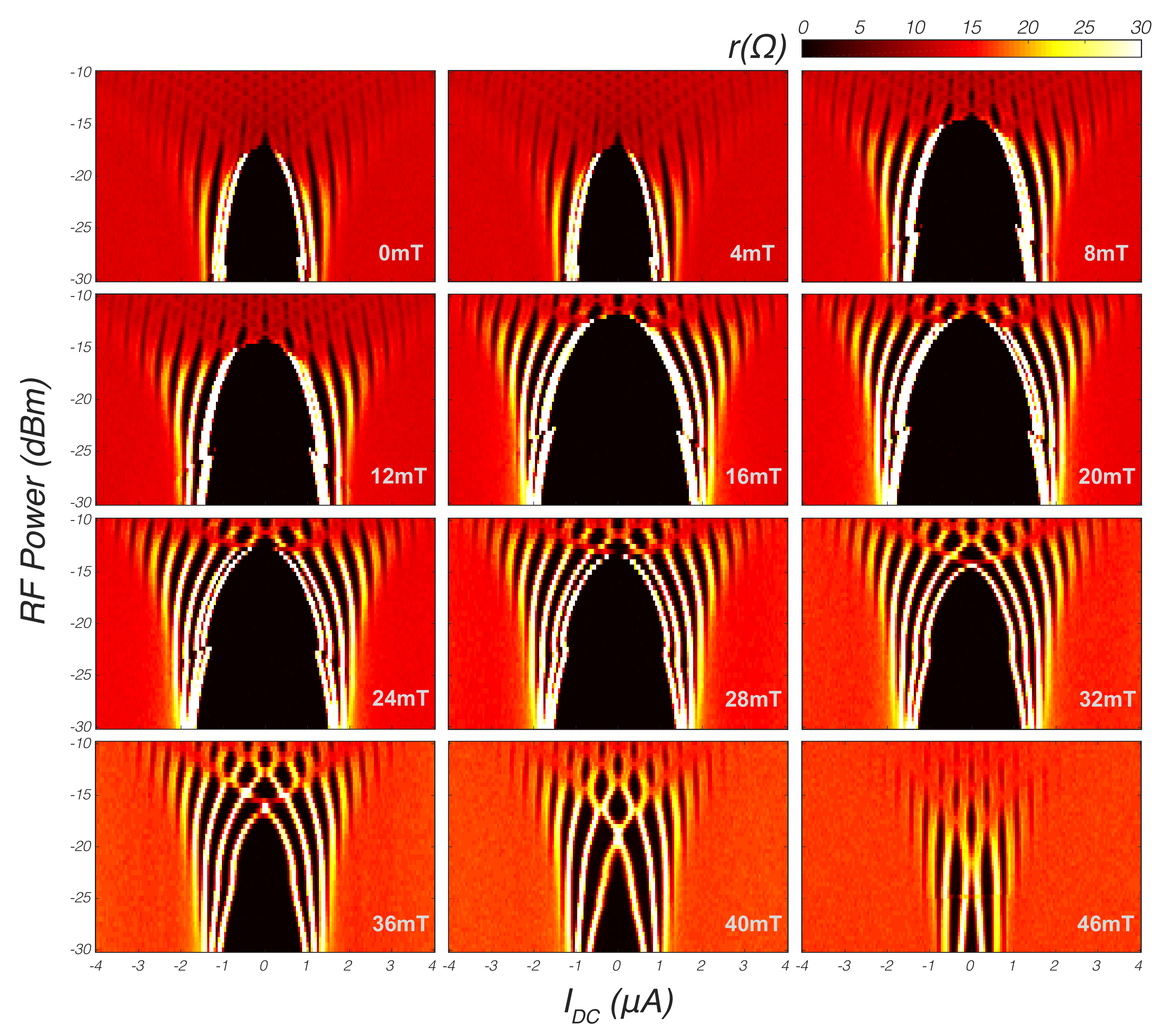}
\caption{\footnotesize{Magnetic field dependence of Shapiro steps for Junction 1, the device highlighted in the main text, at 25 mK.}}
\end{figure*}



\renewcommand{\thefigure}{S6}
\begin{figure}[h]
\center \label{figS6}
\includegraphics[width=\linewidth]{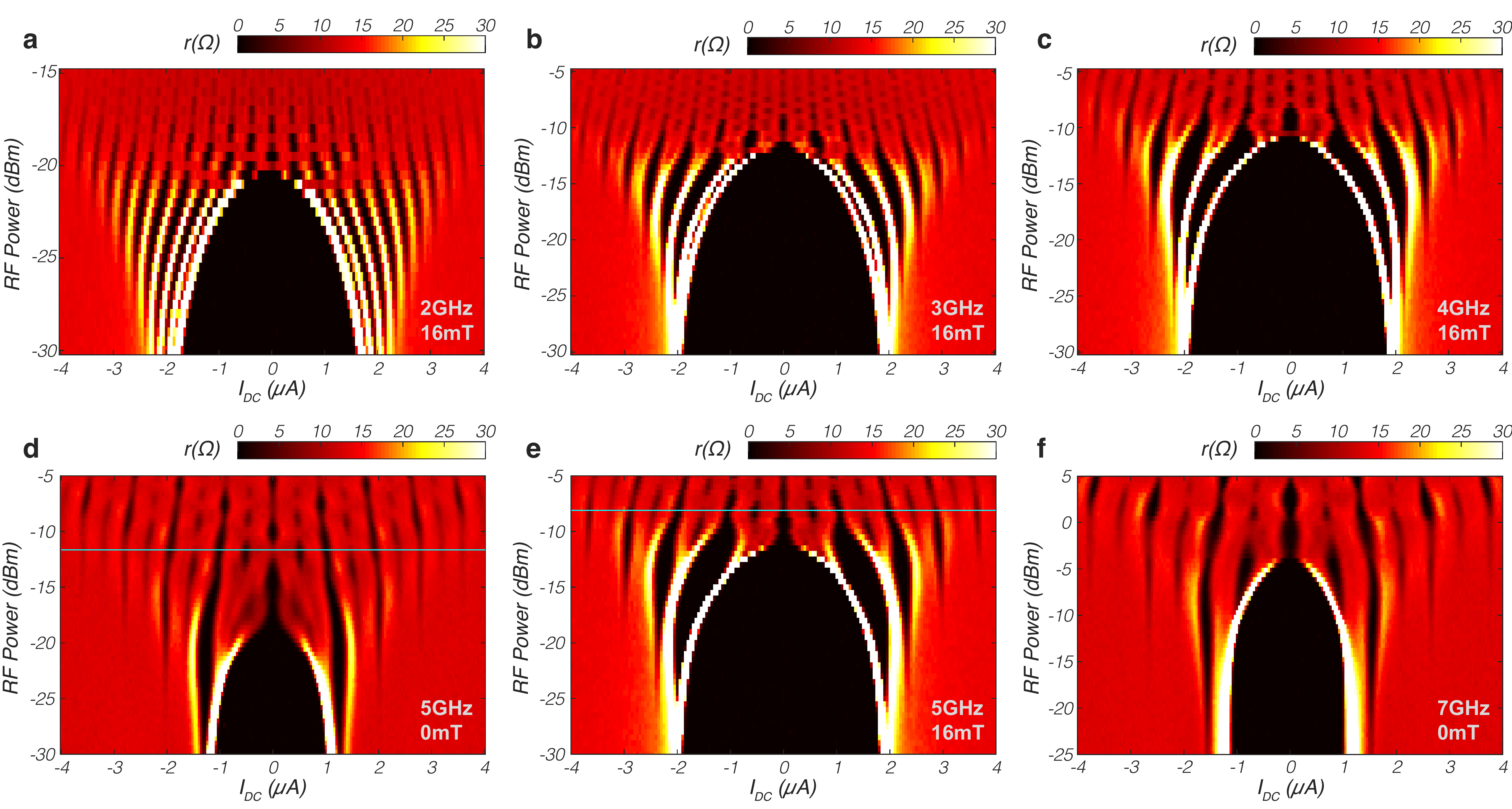}
\caption{\footnotesize{Additional RF data for Junction 1 taken at (\textbf{a}) 2 GHz and 16 mT; (\textbf{b}) 3 GHz and 16 mT; (\textbf{c}) 4 GHz and 16 mT; (\textbf{d}) 5 GHz and 0 mT; (\textbf{e}) 5 GHz and 16 mT; (\textbf{f}) 7 GHz and 0 mT. The light blue lines in (\textbf{d}) and (\textbf{e}) correspond to the cuts in Figs. 4a and 4b of the main text. }}
\end{figure}

\section{Temperature Dependence}

Fig. S7a shows the critical current of the featured junction vs $B$ for various measurement temperatures. As the fridge temperature increases, the prominence of the $B$ = 0 minimum is lessened. At high temperatures ($>$500 mK), this feature is washed out.

Meanwhile, the Shapiro steps and subharmonic features observed in the device are robust against temperature change, as shown in Fig. S7b,c. Although the prominence of the features is decreased as the measurement temperature increases, the same step pattern can still be observed even at temperatures exceeding 900 mK.

\renewcommand{\thefigure}{S7}
\begin{figure}[h]
\center \label{figS7}
\includegraphics[width=\linewidth]{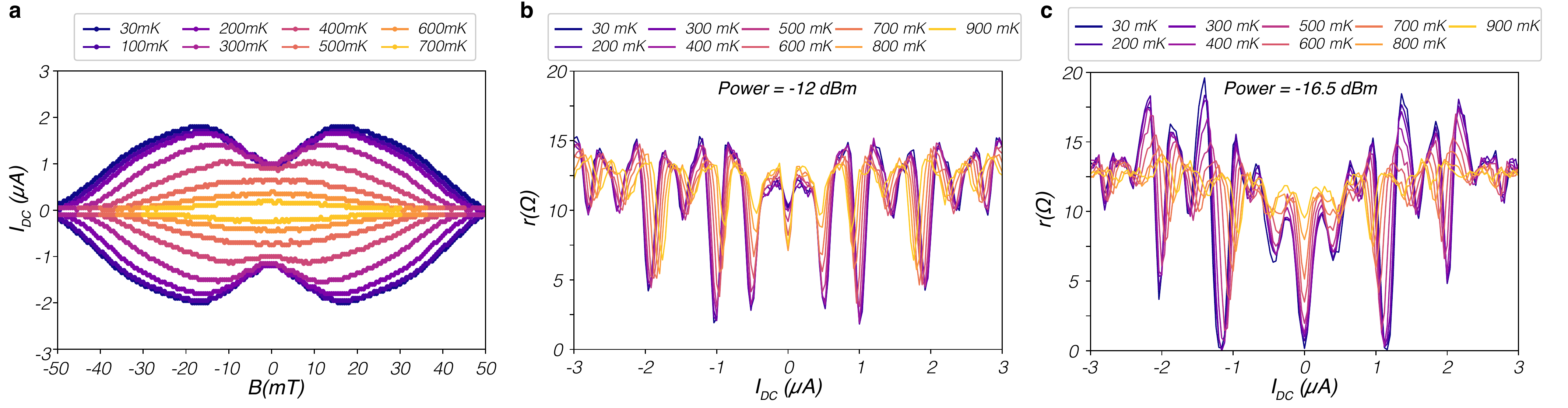}
\caption{\footnotesize{(\textbf{a}) Critical current response to magnetic field applied to Junction 1 at various measurement temperatures. (\textbf{b})  Temperature dependence of the Shapiro step pattern at 5 GHz, -12.0 dBm, 0 mT. (\textbf{c})  Temperature dependence of the pattern at 5 GHz, -16.5 dBm, 16 mT.}}
\end{figure}

\section{Magnetic Diffraction Pattern}

The magnetic diffraction pattern ($r(I_{DC}, B)$) is shown in Fig. S8a, where $B$ is applied perpendicular to the sample substrate. Unlike the MDPs of typical Josephson junctions~Ref. [S1], SnTe junctions display a local minimum of the critical current at zero magnetic field. The peak in $I_C$ occurs at $B$=16mT which, when using the area of the junction (defined as the length of the junction plus twice the penetration depth), corresponds to a flux through the device of $\sim \Phi_0/4$ (where $\Phi_0$ is the quantum of flux). This contrasts with the Fraunhofer-resembling patterns that have been observed in junctions with weak links of bulk TCIs~Ref. [S2],  topological insulators~Ref. [S3-5], and strong-spin-orbit 1D wires~Ref. [S6], where a maximum in $I_C$ at $B$=0 is still observed.  The patterns more closely resemble diffraction patterns for superconductor-ferromagnetic-superconductor~Ref. [S7] and d-wave domain wall~Ref. [S8] Josephson junctions. Measurements in a parallel field do not produce this effect (see section below), ruling out spin-orbit or phase-coherent effects being the origin of the rise in $I_C$ away from $B=0$.

We have observed a total of six different Josephson devices with anomalous magnetic diffraction patterns similar to the one presented in the letter. Table S1 gives the junction length and nanowire diameter as measured by SEM for each of these devices, as well as the field at which the first lobe of the corresponding magnetic pattern closes. The devices are ordered by smallest to largest junction area; Junction 1 is the device highlighted in the paper. As junction area increases, the field at which the lobe closes decreases, showing a clear correlation between junction dimensions and anomalous behavior.

\begin{table*}
\centering
\begin{tabular}{ c |  c  c  c }

		& \shortstack{Nanowire Diameter\\ (nm)}	& \shortstack{Junction Length\\ (nm)} 	&\shortstack{ First Lobe Closes\\ (mT)} \\  \hline 
Junction 1 	& 160			     				& 120 						& 53 \\
Junction 2 	& 220 							& 120 						& 57 \\
Junction 3 	& 230 							& 110 						& 59 \\
Junction 4 	& 280 							& 150 						& 37 \\
Junction 5 	& 390 							& 170 						& 15 \\
Junction 6	& 450 							& 140 						& 12 \\ [1ex]

\end{tabular}
\caption{\footnotesize{Approximate junction dimensions as determined via SEM, and the approximate fields $B$ at which the first lobe of the corresponding magnetic diffraction pattern closes.}}
\end{table*}

While magnetic patterns were collected for all of the devices mentioned in the table, RF lines were connected for Junctions 1 and 4 only, with these measurements yielding similar results. Thus, the remainder of this section will showcase data from Junction 4.

The magnetic diffraction pattern for Junction 4 is shown in Fig. S8d. Rather than an intrinsic effect like the one seen in Junction 1, the considerable hysteresis at positive bias currents can be attributed to self heating of the junction. Indeed, this feature can be confirmed as a situational artifact by sweeping $\Idc$ in opposite directions (from positive to negative bias and vice versa); in this case we observe a mirroring of the data across the $\Idc$ = 0 axis. Furthermore, the effect of this asymmetry on our measurements of the AC Josephson effect, discussed below, qualitatively agree with similar measurements attributed to electron overheating~(Ref. [S9]). 

RF data collected for Junction 4 is presented in Fig. S9, showing a clear magnetic field dependence. This data is qualitatively very similar to that observed from Junction 1, showing deep half steps and additional 1/3 steps. Note that the maximum in $\Ic$ for this device occurs at 6.5 mT; similarly to the data from Junction 1, one can see that the half steps are most prominent at this field.

\renewcommand{\thefigure}{S8}
\begin{figure}[h]
\center \label{figS8}
\includegraphics[width=\linewidth]{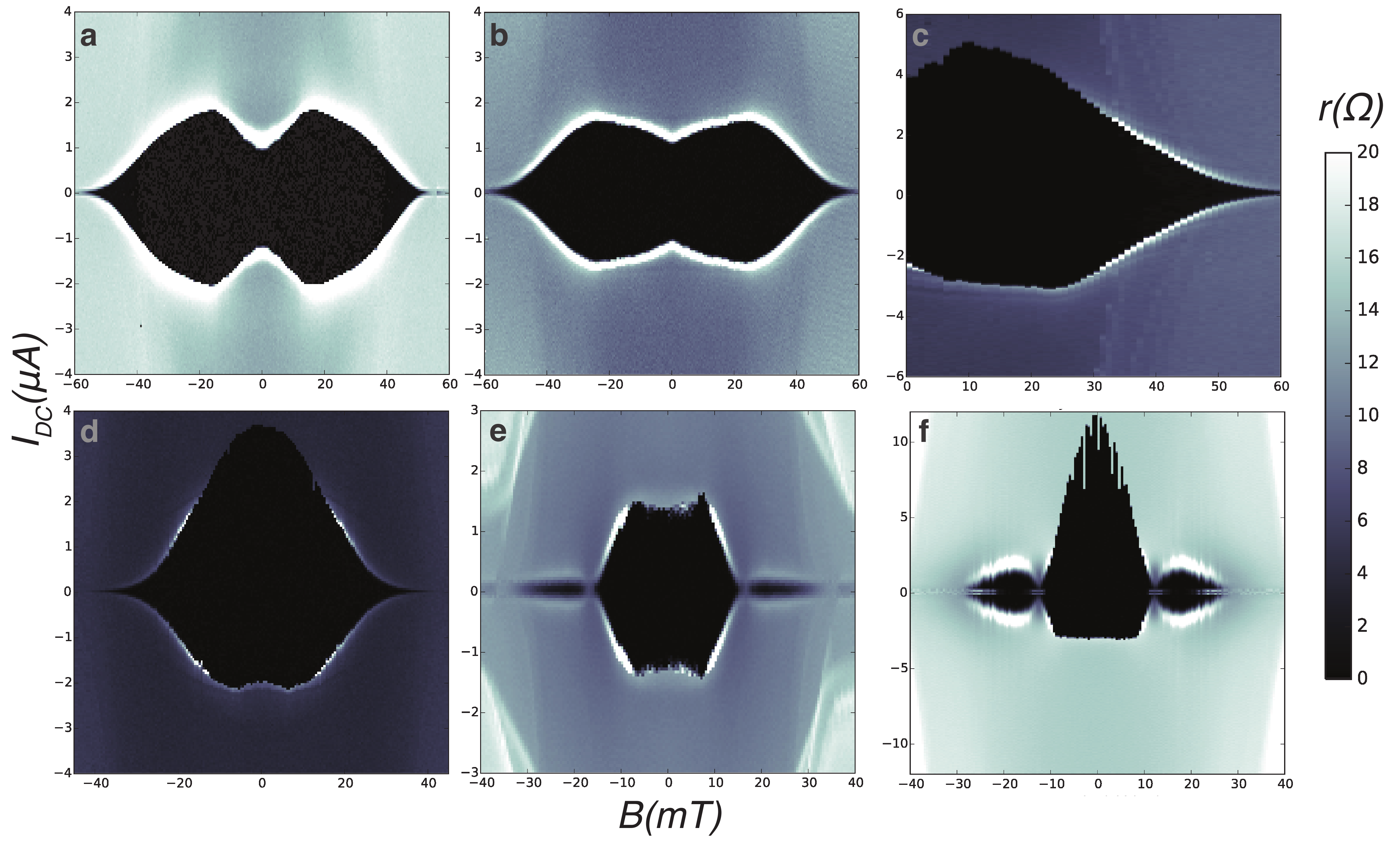}
\caption{\footnotesize{\textbf{a-f.} Magnetic diffraction pattern the junctions in Table I. Junctions 1-6 correspond to Figs. S8 (a-f) respectively. }}
\end{figure}

\renewcommand{\thefigure}{S9}
\begin{figure*}[t]
\center \label{figS9}
\includegraphics[width=\linewidth]{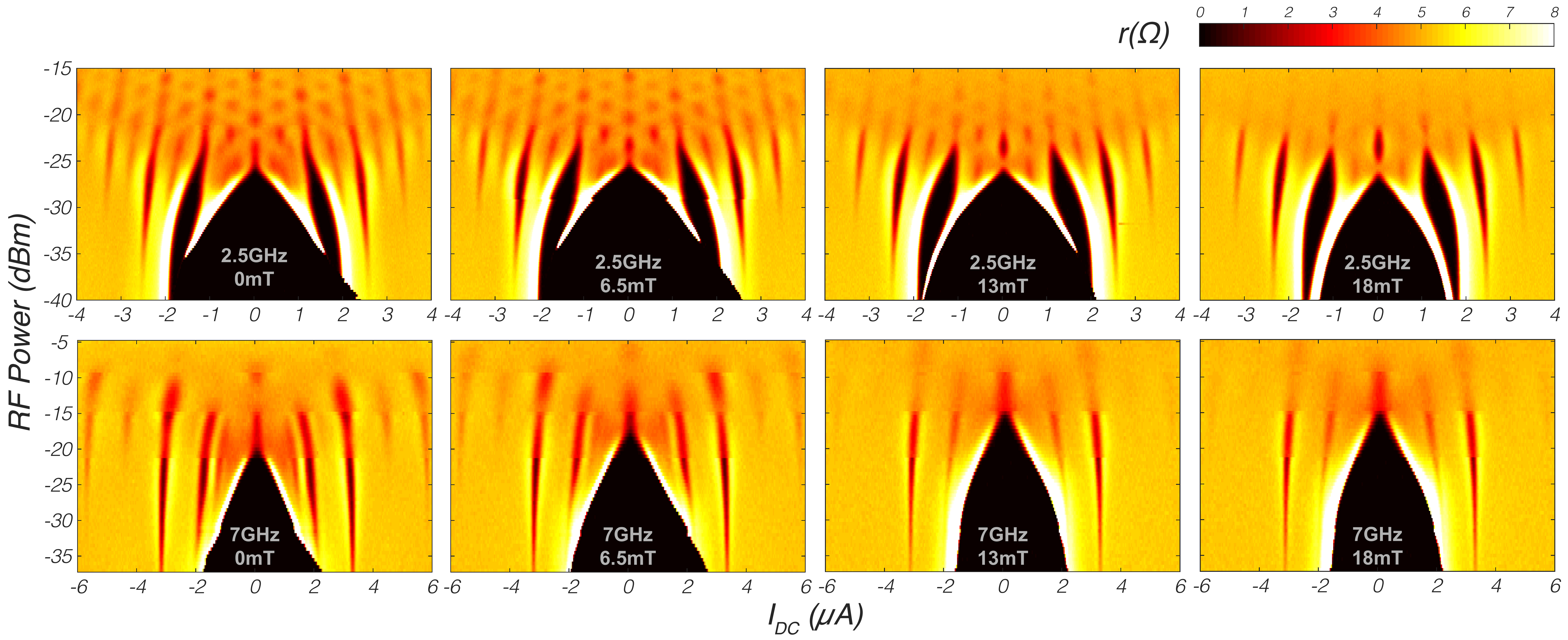}
\caption{\footnotesize{Magnetic field dependence of Shapiro steps for Junction 4 at 2.5 and 7 GHz..}}
\end{figure*}

\section{Magnetic Diffraction Patterns in a Parallel Field}

We place Junction 1 in an in-plane magnetic field perpendicular to the axis of the nanowire in a different cooldown. As shown in Fig. S10, the magnetic diffraction pattern shows a maximum $I_C$ at $B_{\parallel}=0$ with a strong hysteresis and switching effect. Thus, the origin of the minimum $I_C$ highlighted in the main text is not from the spin-orbit or phase-coherent effects.

We also measure a SnTe nanowire contacted by gold leads in a four terminal measurement with $B_{\parallel}$ pointing along the nanowire. The result, an unchanging normal state resistance for the wire, is presented in Fig. S11.

\renewcommand{\thefigure}{S10}
\begin{figure}[h]
\center \label{figS10}
\includegraphics[width=\linewidth]{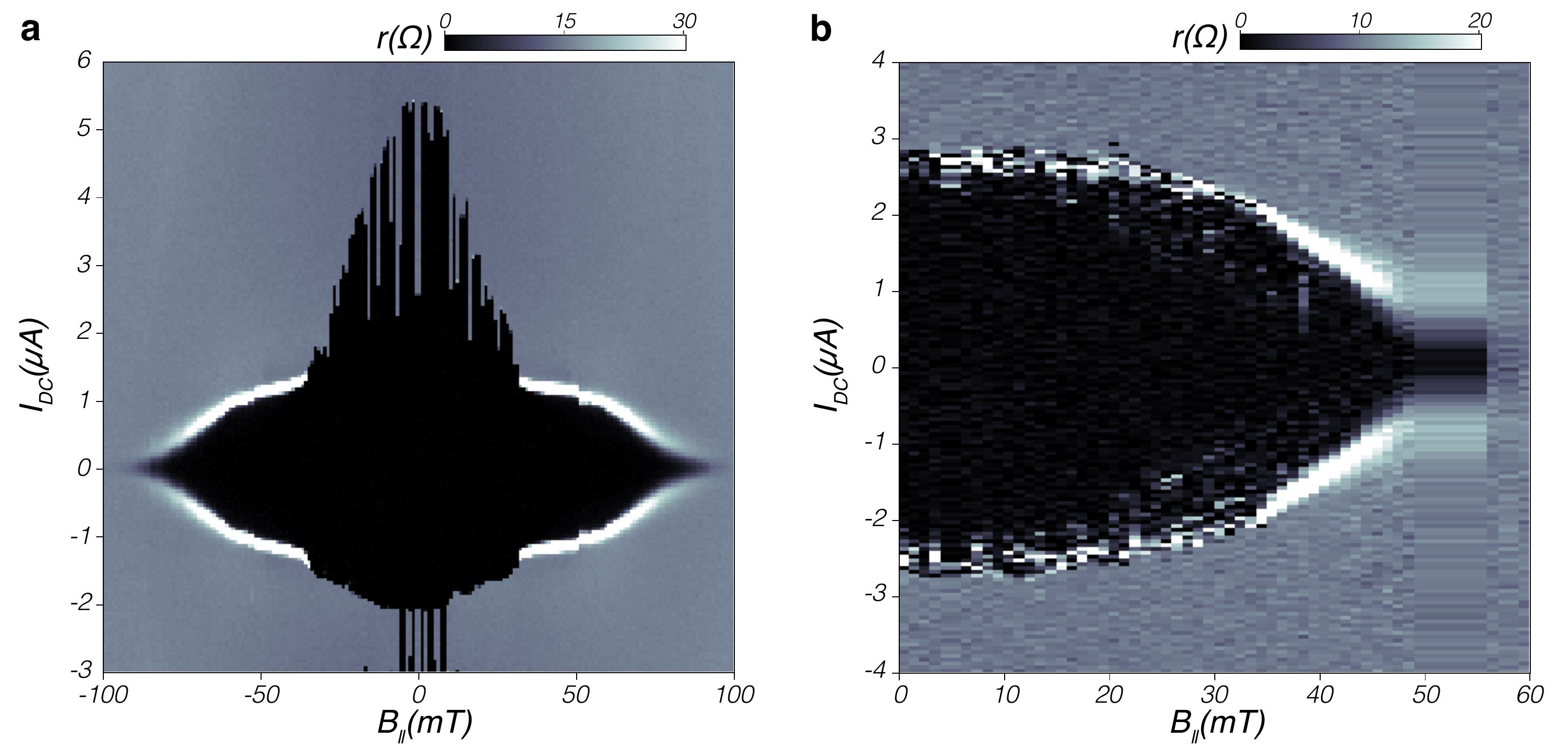}
\caption{\footnotesize{\textbf{a}, Magnetic diffraction pattern for Junction 1 when an in-plane magnetic field $B_{\parallel}$ is applied perpendicular to the axis of the nanowire. The in-plane field data shows a regular maximum of $I_C$ at $B_{\parallel}=0$ with the lobe closes further at about $\pm 90mT$. \textbf{b}, Magnetic diffraction pattern for Junction 2 as a function of $B_{\parallel}$ also shows a maximum at $B_{\parallel}$=0. }}
\end{figure}

\renewcommand{\thefigure}{S11}
\begin{figure}[h]
\center \label{figS11}
\includegraphics[width=0.8\linewidth]{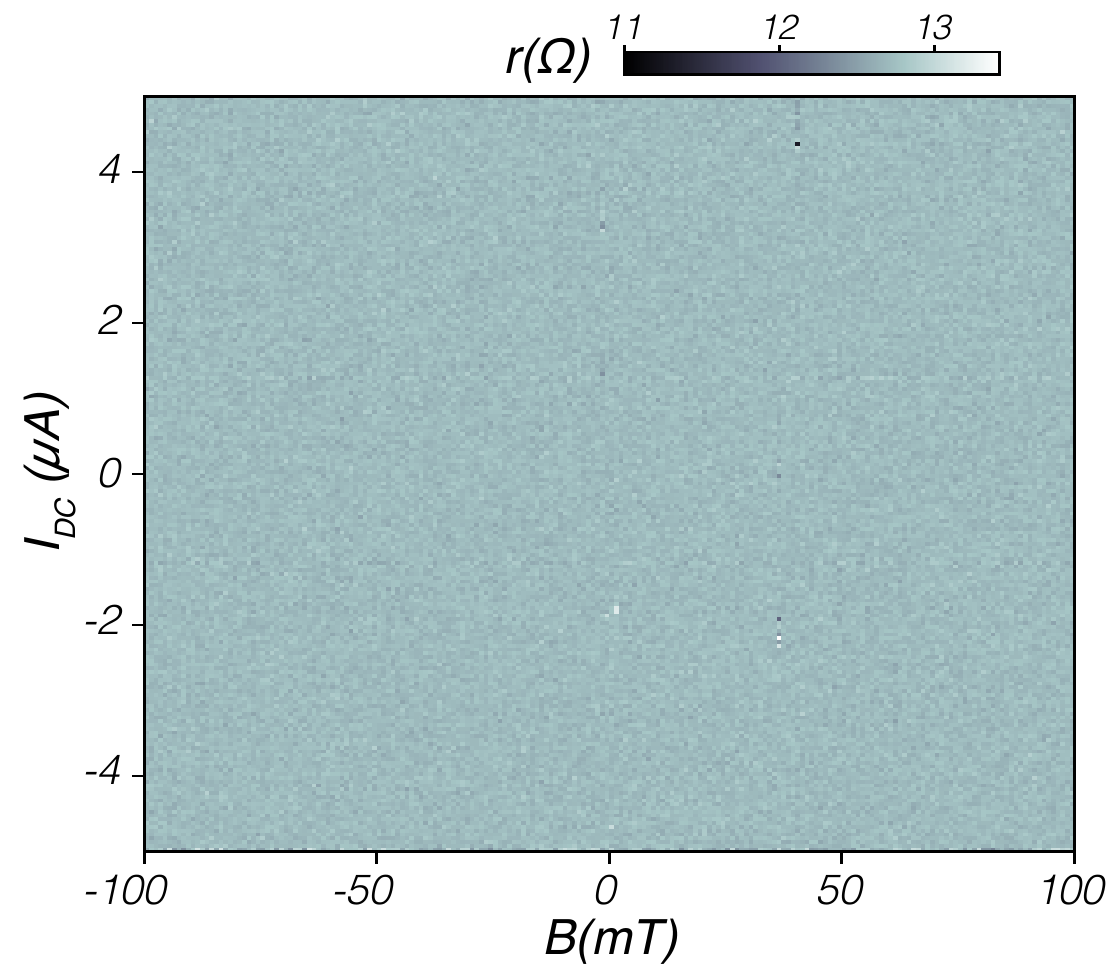}
\caption{\footnotesize{Magnetic diffraction pattern for a SnTe wire contacted by gold leads which shows a constant normal state resistance.}}
\end{figure}


\section{DC Simulation with Resistively-Shunted Junction Model}

The sweeps of DC current bias $I_{DC}$ in two opposite directions in Fig. 3a show two different critical currents $I_C^{+}$ and $I_C^{-}$. The fact that there is no hysteresis suggests that the junction is overdamped. Therefore, we can model the phase difference evolution under a current-phase relation $I(\phi)$ by the resistively shunted junction (RSJ) model:

\begin{equation}\label{eq:RSJ}
 \frac{d\phi}{dt}=\frac{2e R_N}{\hbar}(I_{DC}-I(\phi)), 
\end{equation}
where $R_N$ is the normal resistance of the junction $\sim 13\Omega$. The CPR contains the second harmonic term $\sin 2\phi$ due to the presence of the half Shapiro steps.
\begin{equation}\label{eq:CPR1}
 I(\phi)=I_C(\sin\phi+A\sin 2\phi),
\end{equation}
where A is defined as the relative amplitude of the second harmonic term. At equilibrium, the voltage across the junction is given by

\begin{equation}\label{eq:Veq}
 V=\frac{\hbar}{2e}\left<\frac{d\phi}{dt}\right>=R_N (I_{DC}-I(\phi)), 
\end{equation} 
To examine the symmetry breaking of $V$ under $I_{DC} \rightarrow -I_{DC}$. We can first see how the following symmetries of $I(\phi)$ could affect the invariance of $V$:

\begin{enumerate}
\item{\textbf{Inversion symmetry}: If $I(-\phi) = -I(\phi)$, $V$ is invariant when $\phi \rightarrow -\phi$ and $I_{DC} \rightarrow -I_{DC}$. }
\item{\textbf{$\pi$-Translation symmetry}: If $I(\phi+\pi)=I(\phi)$, $V$ is invariant under $\phi \rightarrow \phi+\pi$ and $I_{DC} \rightarrow -I_{DC}$.}
\end{enumerate}
Thus, the source-drain asymmetry occurs, i.e., $V$ is no longer invariant under $I_{DC} \rightarrow -I_{DC}$, when both symmetries are broken. The CPR in Eq. \ref{eq:CPR1} is inversion but not $\pi$-translation symmetric. By engineering an extra phase term $\beta$, which is not a multiple of $\pi$, in the both terms, we could not only produce a nonzero supercurrent at $\phi=0$ as described in the main text but also model this symmetry breaking by computing the RSJ model.

\begin{equation}\label{eq:CPR2}
I(\phi)=I_{C}(\sin (\phi) +A \sin (2\phi+\beta)
\end{equation}
First, we define the time steps from 0 second to $300/f_C$ with $dt=0.01/f_C$, where $f_C=2eI_C R_N/h$ is the characteristic frequency. Then $\phi(t)$ can be solved the following equation by the \texttt{odeint} function in Python. 

\begin{equation}\label{eq:simulation}
\frac{d\phi}{dt}=\frac{2e R_N}{\hbar}\left\{ I_{DC}-I_{C} \left[ \sin(\phi)+A\sin(2\phi+\beta) \right] \right\}, 
\end{equation}

where we use the parameters $R_N=13\Omega$, $I_{C}=2\mu A$. The relative amplitude $A$ is extracted by comparing the depth ratio of the $N=1/2$ and the $N=1$ Shapiro step in Fig. 4b, which gives about $A \approx 0.909$. $I_{DC}$ is swept from $-3I_C$ to $+3I_C$ and the fitting parameter $\beta$ is swept from $-\pi$ to $\pi$. Then the voltage across the junction for each value of $I_{DC}$ and $\beta$ can be obtained by averaging the last 15 periods ($\Delta t=15/f_C$).
The differential resistance at each $I_{DC}$ can then be calculated by averaging the neighboring voltage difference divided by the current step. The simulation result is presented in Fig. 3b. The ratio $I_{C+}/I_{C-}$ extracted at $B=0$mT in Fig. 3a is about 0.83. By comparing the ratio $I_{C+}/I_{C-}$ at each value in Fig. 3b, the fitting parameter $\beta$ is about $(0.16,0.84)\pi$. Similarly, in Fig. 3a, we could extract the ratio for each field, then map it onto the corresponding $\beta$, as shown in Fig. 3c.

\section{Shapiro Diagram Simulation with Resistively-Shunted Junction Model}

For the Shapiro diagrams, we use the current-phase relation of Eq. \ref{eq:CPR2} and add the RF radiation term in the RSJ model in Eq. \ref{eq:simulation2}:

\begin{equation}\label{eq:simulation2}
\frac{d\phi}{dt}=\frac{2e R_N}{\hbar}\left[ I_{DC}+I_{RF}\sin(2\pi f_{RF}t)-I_{C}(\sin(\phi)+A\sin(2\phi+\beta)) \right], 
\end{equation}

An array of RF power $P_{RF}$ is selected logarithmically from about -85 to -45 dBm then converted into a linear array of $I_{RF}$ by the equation of $I_{RF} = \sqrt{10^{(P_{RF}/10)/Rn/10^3)}}$. Here, we define the time steps from 0 second to $300/f_{RF}$ with $\Delta t=0.02/f_{RF}$, where we choose $f_{RF}=1.1f_C \approx $ 13.9GHz as the driving frequency. For each $P_{RF}$, the dc bias is swept from $-5I_C$ to $5I_C$. At each grid of $(I_{DC}, P_{RF})$, the phase $\phi(t)$ is numerically calculated by $\phi(t+\Delta t)=\phi(t)+\frac{d\phi(t)}{dt}\Delta t $ with the aid of Eq. \ref{eq:simulation2}. We take the mean of the gradients of $\phi(t)$ with respect to $t$ over the last 30 periods ($270/f_{RF}<t <300/f_{RF}$) to obtain the corresponding Josephson voltages  $V=\frac{\hbar}{2e}\left<\frac{d\phi}{dt}\right>$. The differential resistance is then calculated by taking the gradients of voltages with respect to the dc bias current.

Fig. S12 shows the results of the simulation. In the presence of the second harmonic term (Fig. S12b), half steps appear that are not present in the regular sinusoidal CPR (Fig. S12a).By introducing a finite phase shift $\beta=0.16\pi$ and $0.84\pi$, the symmetry of $V$ under $I_{DC}\rightarrow -I_{DC}$ is broken in the low power regimes (Fig. S12 c and d). In the experiment, the half steps next to the 0th step is merged and the asymmetry only reveals at the lowest powers (similar to the breakdown in dc Josephson effect discussed in the main text).

\renewcommand{\thefigure}{S12}
\begin{figure}[h]
\center \label{figS12}
\includegraphics[width=\linewidth]{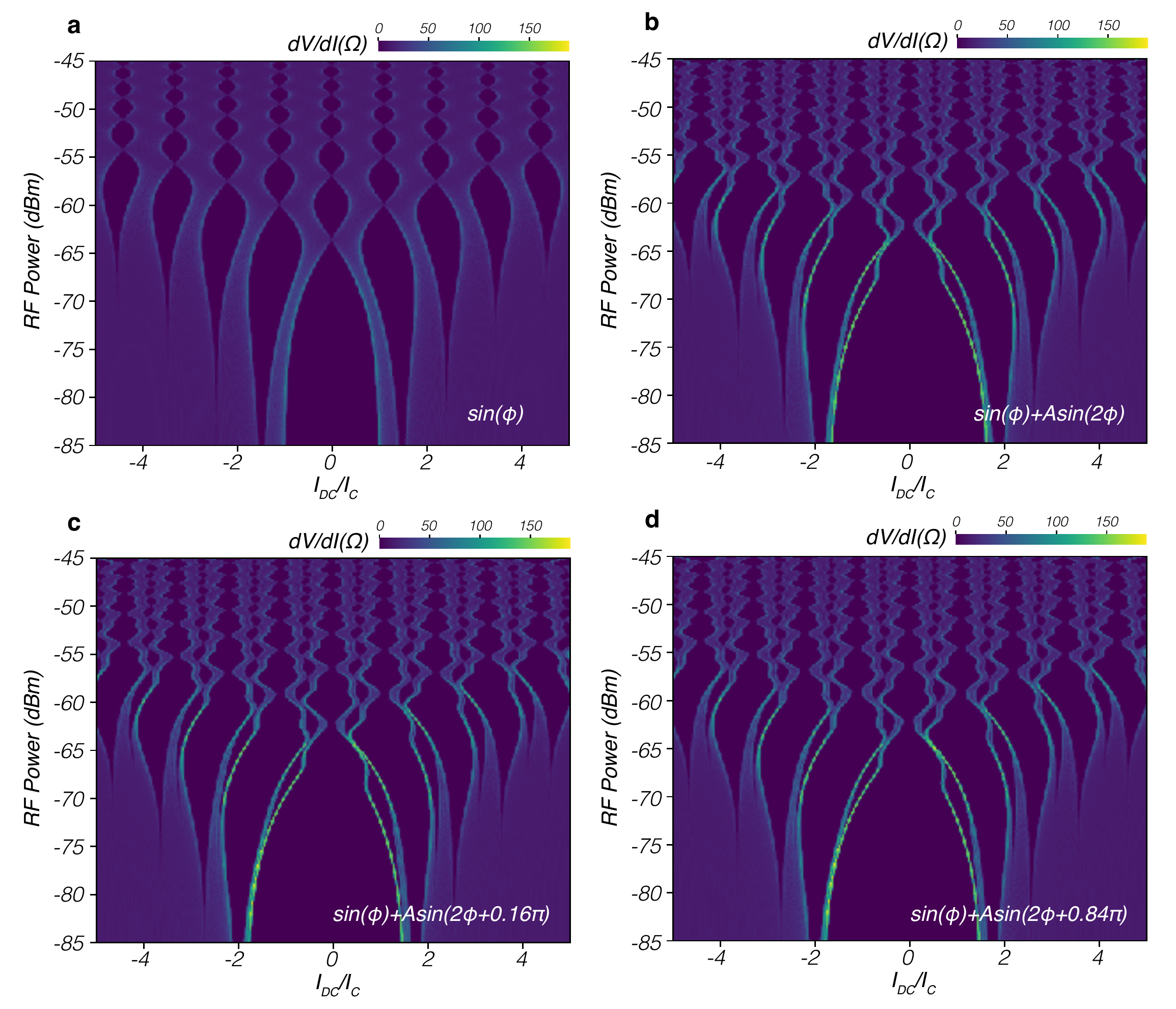}
\caption{\footnotesize{Simulated Shapiro diagrams at $f_{RF}/f_C=1.1$ with Resistively Shunted Junction Model. \textbf{a}, regular sinusoidal CPR: $I(\phi)=\sin(\phi)$. \textbf{b}, CPR with zero phase shift in the second harmonic term: $I(\phi)=\sin(\phi)+A\sin(2\phi)$. \textbf{c} and \textbf{d}, CPR with a finite phase shift in the second harmonic term: $I(\phi)=\sin(\phi)+A\sin(2\phi+\beta)$, where $\beta = 0.16\pi$ and $0.84\pi$, respectively. }}
\end{figure}

\section{\emph{In-situ} cryo-TEM experiments}
SnTe nanowires measured in the present work were synthesized by metal-catalyzed chemical vapor deposition and their transport properties have been reported in our previous works (Ref. [S10]).  The nanowires were characterized by transmission electron microscopy (TEM) for their atomic structure as well as chemical composition.  A high resolution TEM image shows clearly resolved lattice fringes that reflect the expected cubic structure of SnTe (Fig. S13a) and the energy dispersive X-ray spectrum shows the expected stoichiometry of Sn:Te = 1:1, in agreement with bulk reference samples (Fig. S13b).

\renewcommand{\thefigure}{S13}
\begin{figure}[h]
\center \label{figS13}
\includegraphics[width=0.5\linewidth]{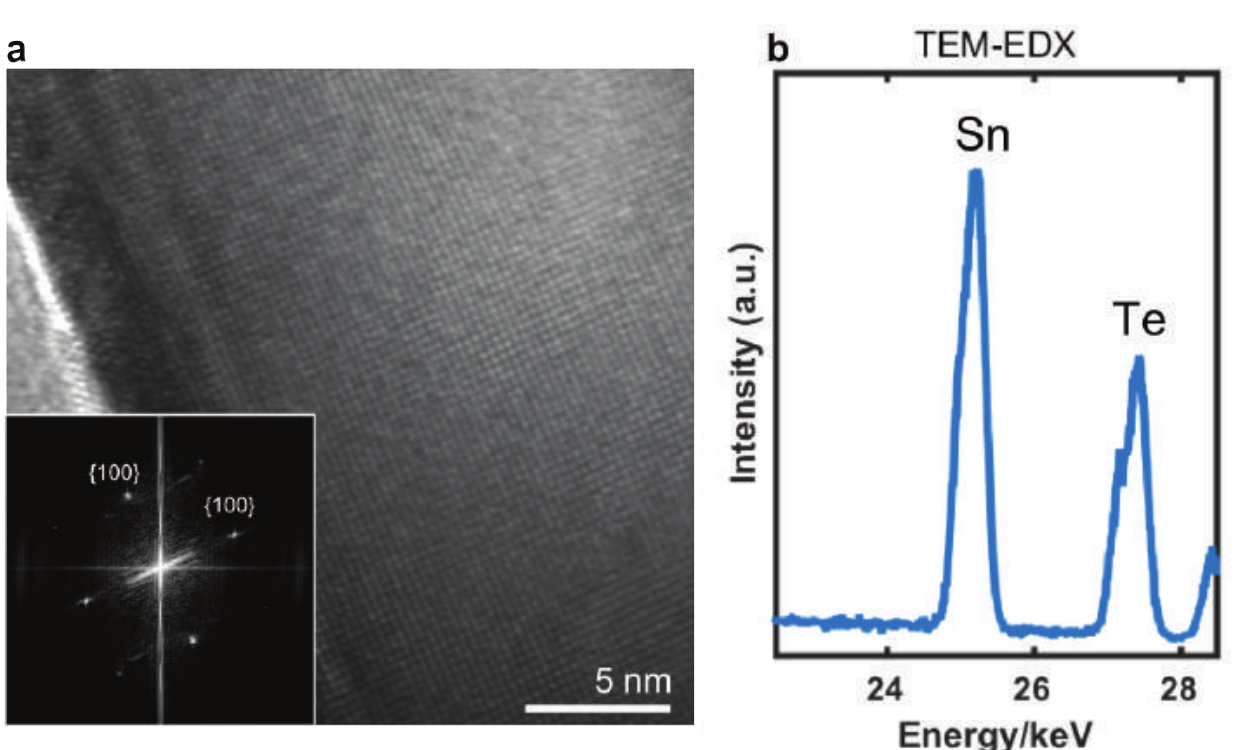}
\caption{\footnotesize{\textbf{a}, High-resolution TEM image shows the expected cubic lattice symmetry from the SnTe nanowire. The inset shows a FFT of the image. \textbf{b}, The stoichiometry of the SnTe nanowire is confirmed using EDX. }}
\end{figure}

When the nanowires were cooled to 12 K in the in situ cryo-TEM experiments, the electron diffraction pattern showed a split of a diffraction spot into two spots, which were separated by 1.2$^o$ (Fig. S14a).  This indicates that the room-temperature cubic phase underwent a phase transformation into a rhombohedral phase with two primary domain directions, as illustrated by the schematics shown in Fig. S14b and c. 

\renewcommand{\thefigure}{S14}
\begin{figure}[h]
\center \label{figS14}
\includegraphics[width=0.75\linewidth]{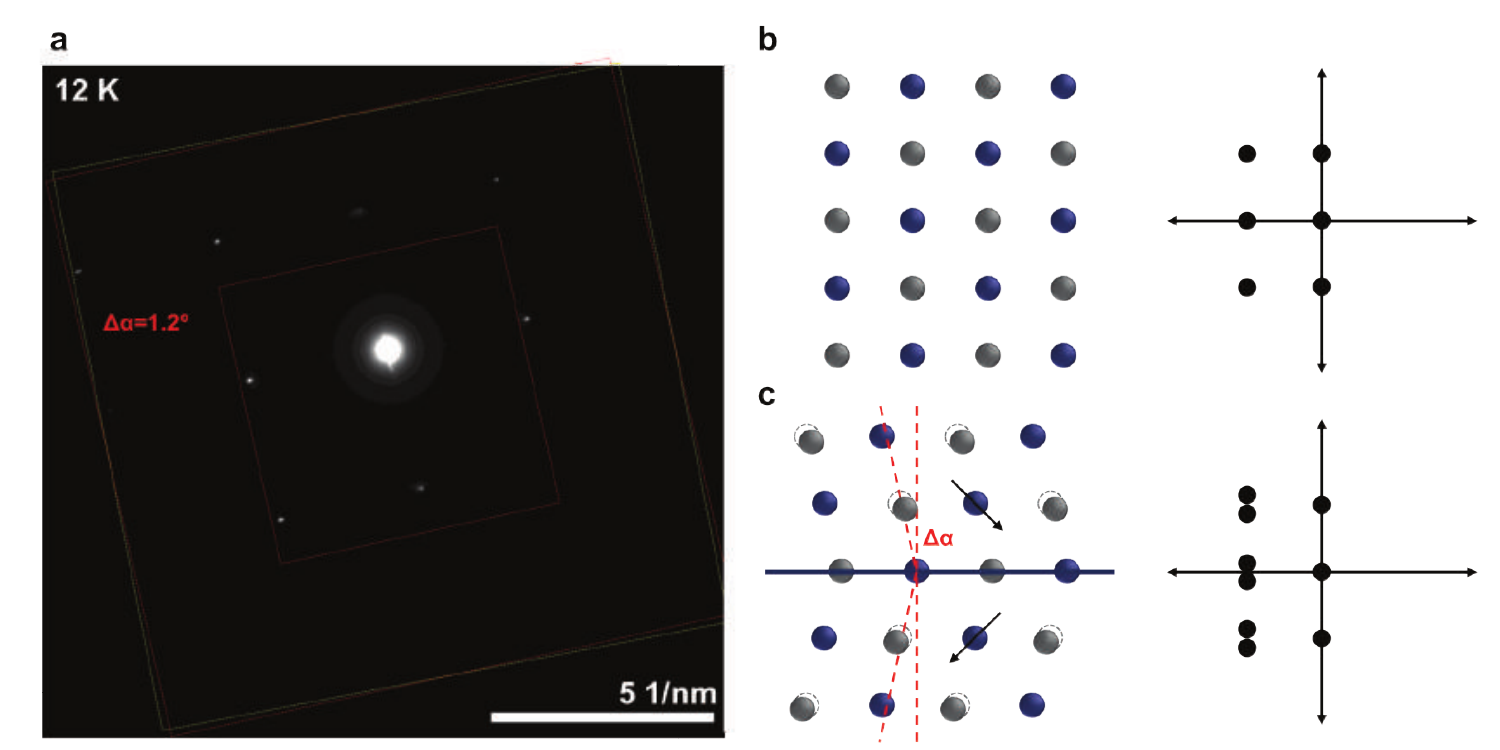}
\caption{\footnotesize{\textbf{a}, Electron diffraction pattern of the SnTe nanowire at 12 K.  \textbf{b}, A schematic of the room-temperature cubic phase (left) and the corresponding electron diffraction pattern (right).  \textbf{c}, A schematic of the low-temperature rhombohedral phase with a domain boundary at the center (left) and the corresponding electron diffraction pattern (right).  The angle between the two rhombohedral domains should be 1.2$^o$, which agrees with the experiment. }}
\end{figure}

Concurrent with the split in the diffraction spot that indicates presence of rhombohedral domains with two primary directions, dark bands appeared along the nanowires at low temperature, which were absent at 290 K (Fig. 2 of the main text).  These dark bands are assigned as domain walls between adjacent rhombohedral (ferroelectric) domains.  This is clearly shown in the in situ movie where the nanowire was gradually warmed up from 12 K to 290 K (supplementary movie S1).  All of the dark bands suddenly disappeared at 80 K, which marks the transition temperature from the low-temperature rhombohedral phase to the high-temperature cubic phase.  Fig. S15 shows TEM images from the in situ movie at various temperatures; dark bands are present for temperatures below 80 K, and absent above 80 K.   

\renewcommand{\thefigure}{S15}
\begin{figure}[h]
\center \label{figS15}
\includegraphics[width=0.5\linewidth]{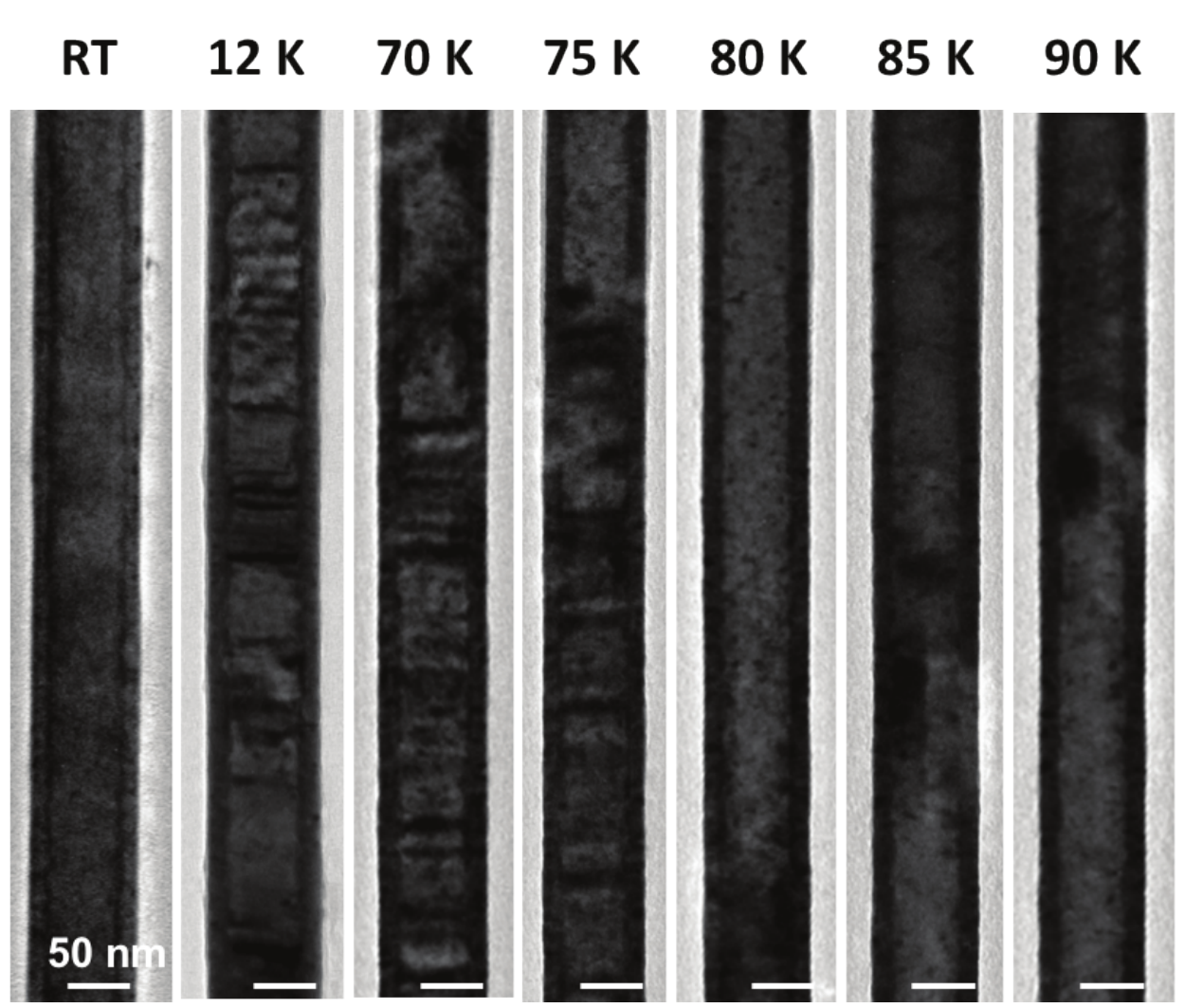}
\caption{\footnotesize{TEM images of a SnTe nanowire warmed up from 12 K.  At the transition temperature of 80 K, all of the dark bands present for temperatures $<$80 K suddenly disappear, clearly indicating that the dark bands mark domain walls between two rhombohedral (ferroelectric) domains.}}
\end{figure}

We also checked that these dark bands were insensitive to the electron beam swing (Supplementary movie S2).  If the positions of the dark bands change as the electron beam is swung, they would be contour bands due to the nanowire not being completely straight.  Fig. S16 shows a series of bright field TEM images of the nanowire at 12 K while the electron beam was swung from right to left and from top to bottom of the field of view.  The dark bands (marked by red arrows) did not move in their positions as the beam was swung.  This proves that the dark bands are not contour bands, and indeed mark the ferroelectric domain walls.

\renewcommand{\thefigure}{S16}
\begin{figure}[h]
\center \label{figS16}
\includegraphics[width=0.5\linewidth]{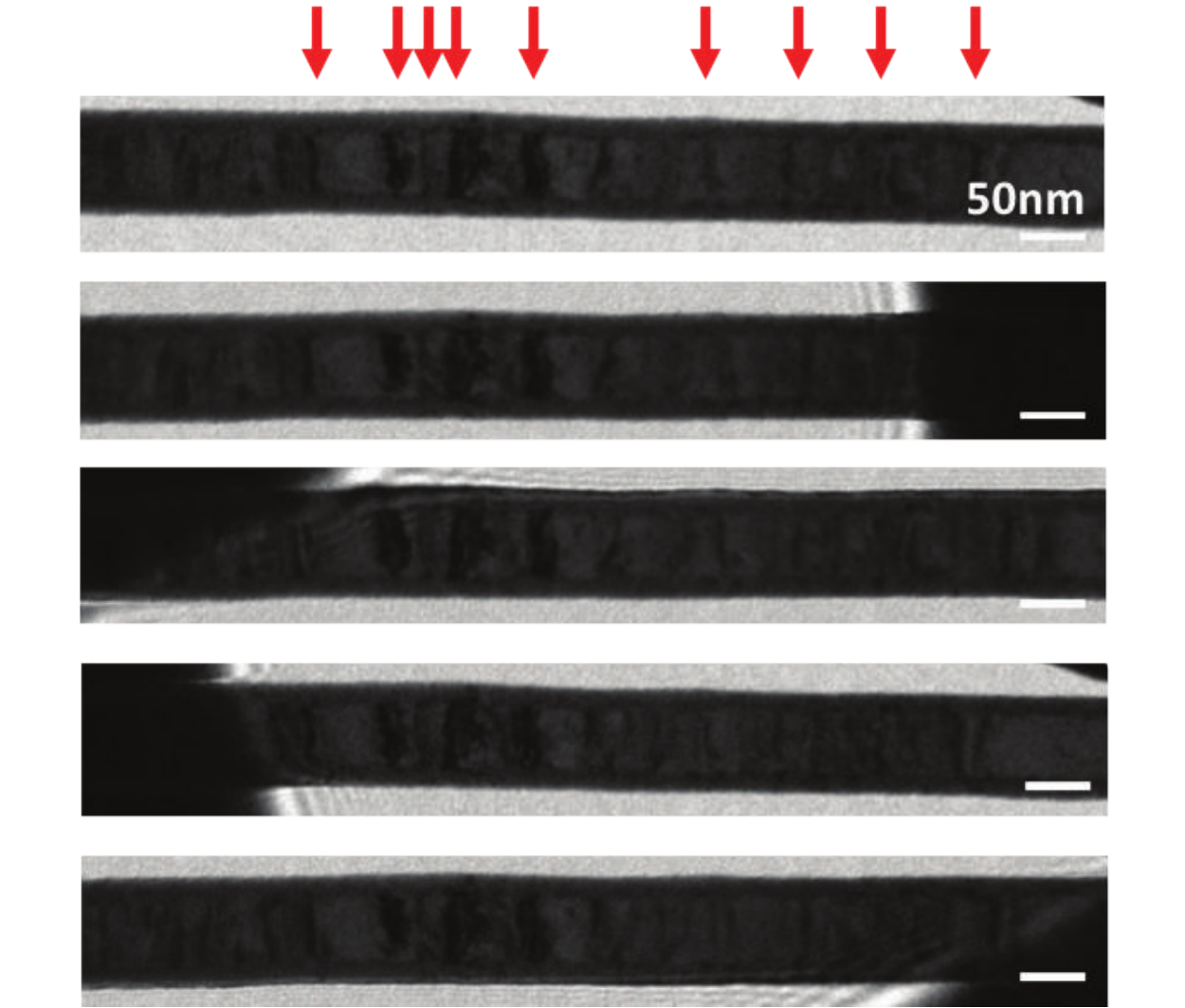}
\caption{\footnotesize{TEM images of a SnTe nanowire with the electron beam swing at 12 K.  As the electron beam was swung, the positions of the dark bands did not change, further confirming that they mark the domain walls between adjacent ferroelectric domains.}}
\end{figure}

A prolonged exposure of the SnTe nanowires to the 200 kV electron beam was found to damage the nanowires during the in situ cryo-TEM experiments.  When the electron beam was focused on the same nanowire for several hours, which was necessary to track the phase transition during warming up, we observed that parts of the nanowires were empty, suggesting severe knock-on damage or sublimation (Fig. S17).  Once the nanowires were damaged, they did not exhibit the expected phase transition.  The nanowire presented in the study was not damaged by the electron beam significantly as the crystallinity of the nanowire at room temperature after the in situ experiment was observed to be similar to the original state at the beginning of the experiment.

\renewcommand{\thefigure}{S17}
\begin{figure}[h]
\center \label{figS17}
\includegraphics[width=0.5\linewidth]{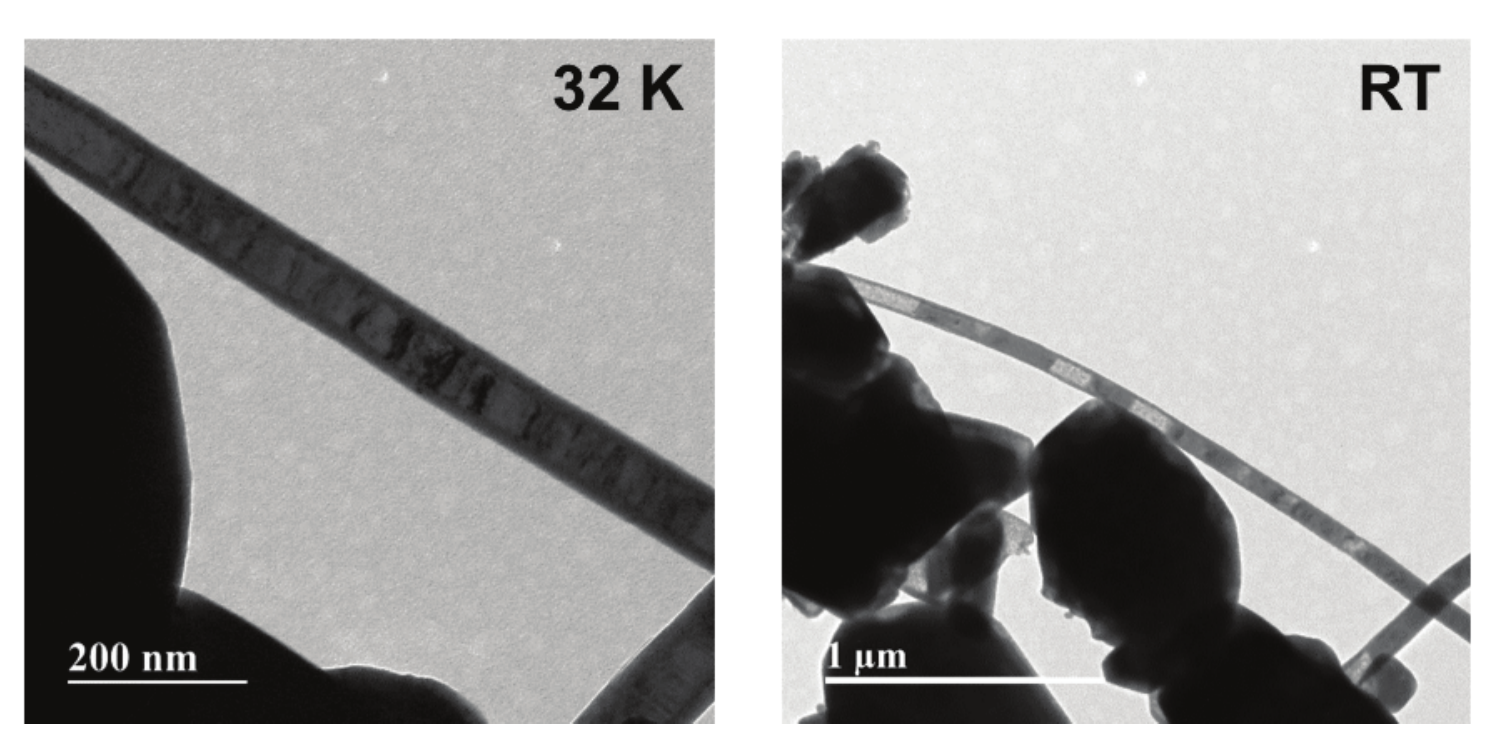}
\caption{\footnotesize{Example of an electron beam damage to SnTe nanowire.  At 32 K, the nanowire shows many dark bands along the nanowire, similar to the nanowire shown in Fig. 2 of the main text.  The nanowire was exposed to the 200 kV electron beam for several hours during the in situ experiment.  Upon a complete warm-up to room temperature, the nanowire was observed to have several voids, which are the brighter regions in the TEM image (image on the right).}}
\end{figure}


\section{Supporting References}

\begin{enumerate}


\item A. Barone, G. Patern\`o, \emph{Physics and Applications of the Josephson Effect}. (Wiley-Interscience Publications, Canada 1982).
\item R. Snyder \emph{et al.},  \emph{Phys. Rev. Lett.} \textbf{121}, 097701 (2018).
\item J. R. Williams \emph{et al.}, \emph{Phys. Rev. Lett.} \textbf{109}, 056803 (2012).
\item M. Veldhorst \emph{et al.}, \emph{Nature Mat.} \textbf{11}, 417 (2012).
\item S. Hart \emph{et al.}, \emph{Nature Phys.} \textbf{10}, 638-643 (2014). 
\item K. Zuo \emph{et al.}, \emph{Phys. Rev. Lett.} \textbf{119}, 187704 (2017).
\item M. Weides \emph{ et al.}, \emph{Phys. Rev. Lett.} \textbf{97}, 247001 (2006).
\item Y. Ishimaru \emph{et al.}, \emph{Phys. Rev. B} \textbf{55}, 11851 (1997). X.-Z. Yan and C.-R. Hu, \emph{Phys. Rev. Lett.} \textbf{83},  1656 (1999).
\item De Cecco, A., Le Calvez, K., Sacépé, B., Winkelmann, C. B., \& Courtois, H. (2016). Interplay between electron overheating and ac Josephson effect. \textit{Physical Review B, 93}(18), 180505.
\item P. Liu \emph{et al.}, \emph{J. of Phys. Chem. Solids} \textbf{128}, 351 (2019), J. Shen \emph{et al.}, \emph{Nano Lett.} \textbf{14}, 4183 (2014), J. Shen \emph{et al.}, \emph{Nano Lett.} \textbf{15}, 4183 (2015).

\end{enumerate}

\end{document}